\documentclass[fleqn,usenatbib]{mnras}

\usepackage{natbib}

\usepackage[T1]{fontenc}

\DeclareRobustCommand{\VAN}[3]{#2}
\let\VANthebibliography\thebibliography
\def\thebibliography{\DeclareRobustCommand{\VAN}[3]{##3}\VANthebibliography}

\usepackage{graphicx}	
\usepackage{amsmath}	
\usepackage{amssymb}	
\usepackage{xspace}
\usepackage{calrsfs}
\usepackage[para,online,flushleft]{threeparttable}
\usepackage{siunitx}
\usepackage[table]{xcolor}
\usepackage{comment}
\usepackage{newtxtext,newtxmath}

\newcommand{\kima}{\texttt{kima}\xspace}
\newcommand{\rebound}{\texttt{rebound}\xspace}

\DeclareMathAlphabet{\pazocal}{OMS}{zplm}{m}{n}
\definecolor{lightgray}{gray}{0.9}
\defcitealias{Doyle2011}{Kepler-16~b}
\defcitealias{Orosz2019}{Kepler-47~}
\defcitealias{Kostov2020}{TOI-1338~b}
\defcitealias{Kostov2014}{Kepler-413~b}
\defcitealias{Orosz2012b}{Kepler-38~b}
\defcitealias{Welsh2012}{Kepler-}
\defcitealias{Socia2020}{Kepler-1661~b}
\defcitealias{Welsh2015}{Kepler-453~b}
\defcitealias{Kostov2016}{Kepler-1647~b}
\defcitealias{Kostov2021}{TIC-1729~b}
\defcitealias{Schwamb2013}{Kepler-64~b}

\title[BEBOP II - Detection limits]{BEBOP II: Sensitivity to sub-Saturn circumbinary planets using radial-velocities}

\author[M. R. Standing et al.]{
Matthew R. Standing,$^{1}$\thanks{E-mail: mxs1263@bham.ac.uk}
Amaury H.~M.~J. Triaud,$^{1}$ 
Jo\~ao P. Faria,$^{2,3}$ 
David V. Martin,$^{4,5}$ 
Isabelle Boisse,$^{6}$ 
\newauthor
Alexandre C.~M. Correia,$^{7,8}$ 
Magali Deleuil,$^{6}$ 
Georgina Dransfield,$^{1}$ 
Micha\"el Gillon,$^{9}$ 
Guillaume H\'ebrard,$^{10}$ 
\newauthor
Coel Hellier,$^{11}$ 
Vedad Kunovac,$^{1}$ 
Pierre F.~L. Maxted,$^{11}$ 
Rosemary Mardling,$^{12}$ 
Alexandre Santerne,$^{6}$\newauthor
Lalitha Sairam,$^{1}$ 
St\'ephane Udry.$^{13}$ 
\\
$^{1}$School of Physics and Astronomy, University of Birmingham, Edgbaston, Birmingham B15 2TT, UK\\
$^{2}$Depto. de F\'isica e Astronomia, Faculdade de Ci\^encias, Universidade do Porto, Rua do Campo Alegre, 4169-007 Porto, Portugal\\
$^{3}$Instituto de Astrofísica e Ci\^encias do Espaço, Universidade do Porto, CAUP, Rua das Estrelas, PT4150-762 Porto, Portugal\\
$^{4}$Department of Astronomy, The Ohio State University, 4055 McPherson Laboratory, Columbus, OH 43210, USA\\
$^{5}$NASA Sagan Fellow\\
$^{6}$Aix Marseille Univ, CNRS, CNES, LAM, Marseille, France\\
$^{7}$CFisUC, Departamento de F\'isica, Universidade de Coimbra, 3004-516 Coimbra, Portugal\\
$^{8}$IMCCE, UMR8028 CNRS, Observatoire de Paris, PSL Universit\'e, 77 av. Denfert-Rochereau, 75014 Paris, France\\
$^{9}$Astrobiology Research Unit, University of Li\`ege, All\'ee du 6 ao\^ut 19 (B5C), 4000 Li\`ege (Sart-Timan), Belgium\\
$^{10}$Institut d'astrophysique de Paris, UMR7095 CNRS, Universit\'e Pierre \& Marie Curie, 98bis boulevard Arago, 75014 Paris, France \\
$^{11}$Astrophysics Group, Keele University, Keele, Staffordshire, ST5 5BG, UK\\
$^{12}$School of Physics and Astronomy, Monash University, Victoria, 3800, Australia\\
$^{13}$Observatoire astronomique de l'universit\'e de Gen\`eve, Chemin Pegasi 51, CH-1290 Versoix, Switzerland\\
}

\date{Accepted XXX. Received YYY; in original form ZZZ}

\pubyear{2021}

\begin{document}
\label{firstpage}
\pagerange{\pageref{firstpage}--\pageref{lastpage}}
\maketitle

\begin{abstract}
BEBOP is a radial-velocity survey that monitors a sample of single-lined eclipsing binaries, in search of circumbinary planets by using high-resolution spectrographs. Here, we describe and test the methods we use to identify planetary signals within the BEBOP data, and establish how we quantify our sensitivity to circumbinary planets by producing detection limits. This process is made easier and more robust by using a diffusive nested sampler. In the process of testing our methods, we notice that contrary to popular wisdom, assuming circular orbits in calculating detection limits for a radial velocity survey provides over-optimistic detection limits by up to $40\%$ in semi-amplitude with implications for all radial-velocity surveys.
We perform example analyses using three BEBOP targets from our Southern HARPS survey. We demonstrate for the first time a repeated ability to reach a residual root mean squared scatter of $3~\rm m\,s^{-1}$ (after removing the binary signal), and find we are sensitive to circumbinary planets with masses down to that of Neptune and Saturn, for orbital periods up to $1000~\rm days$.
\end{abstract}

\begin{keywords}
techniques: radial velocities -- stars: low-mass -- binaries: eclipsing -- software: simulations -- planets and satellites: detection
\end{keywords}



\section{Introduction}\label{sec:intro}

Circumbinary planets are planets which orbit around both stars of a central binary system. Long postulated \cite[e.g.][]{Borucki1984, Schneider1994}, most unambiguous discoveries have only been made in the past decade thanks to the transit method. In total, 14 circumbinary planets have been identified orbiting 12 main-sequence eclipsing binaries. The {\it Kepler} space telescope discovered 12 of these planets orbiting 10 binaries, \cite[]{Doyle2011, Welsh2012, Orosz2012a, Orosz2012b, Schwamb2013, Kostov2013, Kostov2014, Welsh2015, Kostov2016, Orosz2019, Socia2020}, and since two have been discovered in \textit{TESS} photometry: EBLM~J0608-68~b/TOI-1338~b \citep{Kostov2020} and TIC~17290098~b \citep{Kostov2021}.

Despite these successes, circumbinary configurations remain largely elusive on account of their longer orbital periods. \cite{Martin2018} and references therein, show how most circumbinary planets found to date lie just outside of the circumbinary stability limit, as described by \cite{HOLMAN1999}, which incidentally places them often close or within the habitable-zones of their parent stars.

Binary-driven orbital precession means that most circumbinary planets find themselves in transiting configurations only $\approx25\%$ of the time \citep{Martin2017}, while their presence would be visible $100\%$ of the time in radial velocity measurements. With sufficient precision, radial-velocities have the potential to be more efficient at detecting circumbinary planets than the transit method, particularly since the Doppler method is also much less sensitive to orbital inclination and period \citep[e.g.][]{Martin2019}. 

Despite these advantages, no circumbinary planet has been discovered by the Doppler method so far. Only one has been detected in follow-up, a recent recovery of Kepler-16~b by our project (Triaud et al. in prep).  To unlock more information on circumbinary formation mechanisms the demographics of circumbinary planets need to increase, along with accurate physical and orbital parameters. 

To this end, \cite{Konacki2009b} launched a Doppler survey in an attempt to detect circumbinary planets with radial velocities alone, with `The Attempt To Observe Outer-planets In Non-single-stellar Environments' (\textit{TATOOINE}) survey. Unfortunately, TATOOINE was unable to discover any planetary companions during their survey. This is likely due to the type of binary stars that were observed. TATOOINE observed double-lined (SB2) binary star systems.
With spectra from both stars visible within a spectrograph, complex deconvolution methods need to be applied to recover the individual components' velocities with high precision and accuracy. \citet{Konacki2009b} found their best binary target was HD~9939 ($V =7$). Their ten radial velocity measurements on this target from Keck HIRES have uncertainties of $1-4~\rm{m\,s^{-1}}$, but they find a residual root mean squared (RMS) scatter of $7~\rm{m\,s^{-1}}$. This RMS is calculated after the data have been fitted for binary Keplerian models, which we refer to as {\it residual RMS} for the remainder of this paper.
To achieve this level of RMS scatter on an SB2 is impressive, though combining all their data on this target yielded a $99\%$ confidence detection limit of $\sim 1~\rm M_{Jup}$ \citep{Konacki2009b}. Results on other targets typically yield residual RMS in excess of $10~\rm{m\,s^{-1}}$
In the end, \cite{Konacki2010} recommend that single-lined binaries might be the next step forward, a step that we took.

\cite{Martin2019} introduced our Binaries Escorted By Orbiting Planets (BEBOP) survey. BEBOP currently only targets proven single-lined (SB1) eclipsing binary targets, where only the spectrum of the primary star, typically an F or G dwarf, is visible. We note here, that whilst the binary mass ratios of the BEBOP and TATOOINE samples are biased towards low and high values, respectively, \cite{Martin2019b} showed that circumbinary planets exist around all mass ratios with no discernable preference.
Our binary sample was identified while confirming transiting hot Jupiters with the Wide Angle Search for Planets ({\it WASP}) survey \citep[]{Pollacco2006, Triaud2013, Triaud2017}. First we used the CORALIE spectrograph, on the 1.2~m {\it Euler} telescope at La Silla, Chile \citep{Martin2019}, and have now extended the survey to HARPS, at the ESO 3.6~m telescope, also in Chile \citep{Pepe2002b}, as well as with SOPHIE, at the OHP 193~cm telescope in France \citep{Perruchot08}. 

In this paper we detail and test our detection and observational protocols for the BEBOP survey, which have improved since \citet{Martin2019}. To fit our data, we now use a diffusive nested sampler called \kima  \citep{Faria2018}. Contrary to our previous method, we now measure evidence for the presence of a circumbinary planet, accounting for Ockham's razor, and can marginalise our results over as-of-yet undetected planets. With \kima, we investigate our detection sensitivity to circumbinary planets, and we use it to produce robust detection limits. We also test our detection protocol by injecting circumbinary planets into our HARPS data and retrieve them with \kima. We show that, after removing the binary signal, we repeatedly achieve a detection limit for circumbinary planets at masses as low as Neptune's, paving our way to actual detections.

This paper is organised as follows. In Sect.~\ref{sec:Observations} we describe our observing strategy for the BEBOP survey along with statistics of data gathered to date. Second in Sect.~\ref{sec:kima} we provide a brief overview of the \kima diffusive nested sampling package, along with justification of its use within the BEBOP survey. In Sect.~\ref{sec:Methods} the data analysis and simulation methods used are described. In Sect.~\ref{sec:results_dis} we display the results of our detection limit and simulation recovery analysis, and discuss the effect of the results on the BEBOP survey before concluding in Sect.~\ref{sec:conclusion}.

\section{Description of our observational protocol and data collection}\label{sec:Observations}

The current BEBOP sample consists of a total of 113 single-lined eclipsing binaries which do not exhibit strong stellar activity, or the presence of a tertiary star. 54 in the Southern hemisphere, with HARPS, and 59 in the Northern hemisphere, with SOPHIE;  observations still ongoing in each. The Southern sample was established from the EBLM programme, which identified low-mass eclipsing binaries from WASP false-positives \citep{Triaud2013, Triaud2017, Swayne2021}. The data we use in this paper is exclusively from our Southern sample, for which we have now accumulated over 1200 HARPS spectra, an average $\sim 23$ individual radial-velocity measurements per target. 
The southern data are obtained from a preliminary proof-of-concept run (Prog.ID 099.C-0138) and a 78 night programme seeking planetary candidates that commenced between April 2018 and March 2020 (Prog.ID 1101.C-0721). BEBOP has been awarded a further large programme (Prog.ID 106.212H), which began in September 2020 to confirm a number of candidate planetary signals, the results of which will be published in future work.

\subsection{Observational protocol}

All systems are observed as homogeneously in time as is possible, with 1800s exposures, at an average cadence of one measurement every $\approx 6$ nights except when a system is no longer visible. All measurements are taken at airmass $<1.6$. We attempt to cover as much of the yearly visibility as is feasible. All measurements are taken using the {\sc Obj\_AB} mode of HARPS, which places the B fibre on the sky rather than on a simultaneous Fabry-P\'erot calibration lamp, with the A fibre placed on our science target \citep{Pepe2002b}. Whilst the {\sc Obj\_AB} mode prevents us from reaching the most accurate mode of HARPS, this gives us a chance to subtract moonlight contamination from our spectra, which can introduce a secondary spectral component, something our sample is designed to avoid. Since most our systems have V magnitudes between 9 and 12, photon noise rarely reaches down to the $1~\rm m\,s^{-1}$ long-term stability of the instrument \citep{Lovis&Pepe2007, Mayor2009}, removing the need for simultaneous calibration. Calibration of the instrument is done at the start of night using Fabry-P\'erot and Th-Ar calibration lamps as is now standard on HARPS \citep{Coffinet2019}.

The data is analysed on a bi-monthly basis for quality control (for instance verifying whether the residual residual RMS around the binary solution is low enough to allow exoplanet detections). Exoplanet candidate identification are done only once a year. We chose on purpose to perform those candidate searches rarely in order to avoid falling into observer's bias. A homogeneously observed dataset also provides more robust detection limits. On account of the rather long orbital periods expected for circumbinary planets \citep[$50+$ days;][]{Martin2018}, we are only now reaching the ability to confirm exoplanetary candidates.

A similar procedure is performed with SOPHIE, on the Northern sample. It will be more specifically described once SOPHIE data are presented for publication.

\subsection{Data collection reduction, and outlier removal}

Observations are obtained using the HARPS instrument on the ESO 3.6m telescope situated at the La Silla observatory in Chile \citep{Mayor2003}. 
Reduction of the spectroscopic data was carried out by the HARPS pipeline \citep{Lovis&Pepe2007}. A Cross Correlation Function \citep[CCF;][]{Baranne1996, Pepe2002a} is created by comparing the spectra obtained by the spectrograph with a G2 or K5 template mask spectra (depending on the spectral type of the primary star in question). We remind here, that our targets are chosen such that the spectra of the secondary star is too faint to be detected. Typically our primary stars are $>4$ magnitudes brighter than the secondaries. This allows us to safely ignore the secondary CCF and treat each system as a single-star system \citep[spectroscopically speaking;][]{Martin2019}.
The correlation is evaluated at $0.5~\rm{km\,s^{-1}}$ intervals and the CCF indicates with its lowest point the radial velocity at which the mask corresponds most closely with the targets' spectra. The CCF is fit with an inverted Gaussian profile and its mean recorded as the radial-velocity.

Various shape metrics are obtained from the CCF itself, such as the span of the inverse of the mean bisector slope (bisector span) and the Full Width at Half Maximum (FWHM). These are often used as indicators of stellar activity \citep{Queloz2001, Santos2002} but mainly they track the quality of our observation and whether any radial-velocity displacement is caused by a change in the shape of the CCF, or by a translation of the CCF, which is what we are after. 

Eleven systems (9 on HARPS and 2 with SOPHIE) with particularly strong anti-correlation have been dropped from the observing schedule. A anti-correlation between bisector span and RV is likely caused by starspots on the surface of the target creating parasitic signals 
\citep[e.g.][]{Queloz2001}. 

Prior to fitting our radial-velocities we clean any obvious outliers from our data. First, we exclude any observations mistakenly obtained on a star other than our target. This is often seen as a significant difference in the spectrum's signal-to-noise and in the FWHM. Second, we remove any measurement likely affected by the Rossiter-McLaughlin effect \citep[e.g.][]{Triaud2018}. To find which measurements are affected, we fit a Keplerian binary model to the radial-velocities, and from the fit parameters compute when eclipses would happen.

In a third step, we examine the distribution of all bisector span and FWHM measurements of a given system, and exclude measurements that are further from the mean by more than $3\sigma$. We also perform target per target visual inspections of these metrics and flag systems where the bisector span and/or the FWHM appear to show a long term trend.

\subsection{Choice of targets \& observation summary}
For this paper we select three systems from within the full BEBOP South sample. We choose our Southern sample for this exercise as we have obtained more high precision data with a longer baseline than what is currently available in BEBOP North. We select these three from the sample for several reasons. The first is that no planetary or stellar activity signal are currently visible. 

\begin{figure}
	\includegraphics[width=\columnwidth]{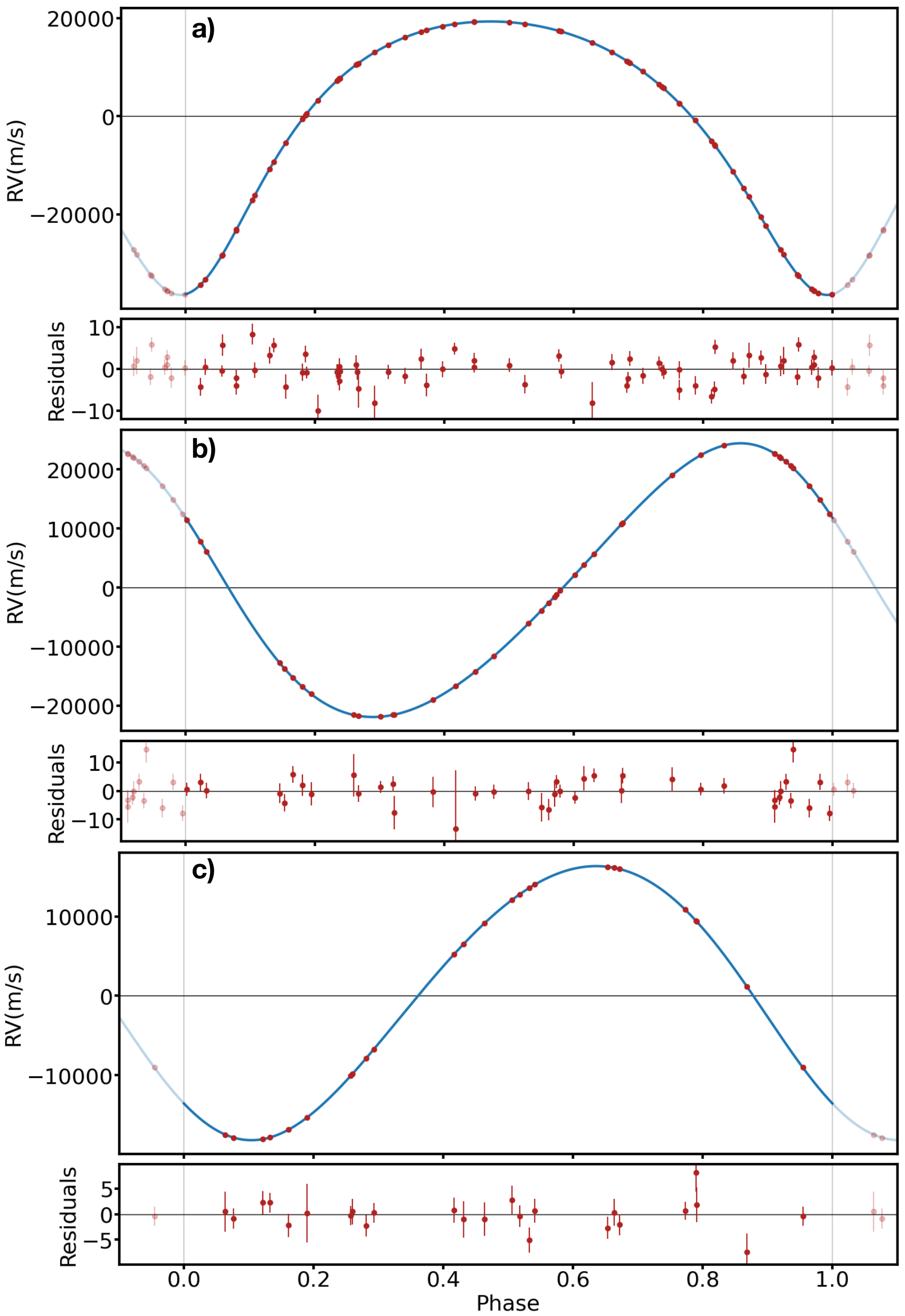}
    \caption{Best fit phased RV models for a) = J0310-31, b) = J1540-09, c) = J1928-38 with associated residuals. Doppler motion of the primaries seen here are caused by the secondary stars. Residuals have an RMS = $3.2$, $3.8$, and $3.0~\rm{m\,s^{-1}}$, respectively.}
    \label{fig:J0310-31_phased_binary}
\end{figure}

Second, we select amongst the systems with the lowest radial-velocity uncertainty, and lowest RMS scatter, after having removed the contribution of the secondary star to the radial-velocities of the primary (that we refer to as {\it residual RMS}). We do this to specifically demonstrate BEBOP's ability to recover circumbinary planets with signals of a few $\rm m\,s^{-1}$.
Of our 41 Southern targets with more than the average of $23$ spectra we have identified $17$ with a residual RMS $< 10~\rm{m\,s^{-1}}$, corresponding to roughly $42\%$ of the sample. While $27$ of these targets ($66\%$) have an RMS  $< 17~\rm{m\,s^{-1}}$.
Three systems are selected from within these, which we now describe:

1) EBLM~J0310-31 (J0310-31 thereafter) has been part of the preliminary survey for BEBOP on HARPS, and of the first extensive observing campaign. This is the target for which the largest number of spectra has been obtained. In total, 65 RV data points are available, obtained between 2017-07-09 and 2020-01-02. These measurements also have the lowest mean uncertainty of the survey $\rm \overline{\sigma_{RV}}~=~2.13~{m\,s^{-1}}$. 
J0310-31's residual $\rm RMS~=~3.16~\rm m\,s^{-1}$, which is one of the smallest we obtain so far. All these make J0310-31 the ideal target to test our procedures to compute detection limits, as well as perform injection-recovery tests on. See figure \ref{fig:J0310-31_phased_binary} for a phased plot of our RV data for J0310-31, along with our best fit model and residuals.

2) EBLM~J1928-38 (J1928-38 for short). Its residual $\rm RMS~=~2.96~ m\,s^{-1}$, better than J0310-31, close to its mean photon noise uncertainty $\rm \overline{\sigma_{RV}}~=~2.67~m\,s^{-1}$. Only $25$ measurements have been obtained on J1928-38 so far, which makes it more representative of the current state of the survey than J0310-31 is. These were collected between 2018-06-04 and 2019-09-14.

3) EBLM~J1540-09 (henceforth J1540-09) has $\rm \overline{\sigma_{RV}}~=~3.86~m\,s^{-1}$, and $\rm RMS~=~3.75~m\,s^{-1}$, for 41 available spectra, observed from 2017-04-20 to 2020-03-06.

These targets were analysed using the \kima RV analysis package (see next section). Their parameters can be found in Table~\ref{tab:System_parameters}. All data can be accessed in Tables~
\ref{tab:J0310-31_data} to Table~\ref{tab:J1928-38_data}. 
All measurements for these three systems are used, with no outliers found within those three datasets.

\begin{table*}
	\centering
	\caption{Updated binary, system and derived parameters for the investigated binary systems. Uncertainties are given in the brackets, as the last two significant figures, except for $\sigma_{\rm jit}$ where uncertainties can be significantly asymmetric.}
	\label{tab:System_parameters}
	\begin{tabular}{llll} 
		\hline
		\hline
		 & EBLM J0310-31 & EBLM J1540-09 & EBLM J1928-38\\
		\hline
		{\it System properties} & & &\\
		TIC & 89045042 & 32431480 & 469755925\\
        TYC & 7019-784-1 & 5600-377-1 & 7931-842-1\\
		Gaia DR2 & 5057983496155992448 & 6317098582556256000 & 6739146911148825344\\
		$\alpha$ [deg]$^1$& 03:10:22.62 & 15:40:08.99 & 19:28:58.85\\
		$\delta$ [deg]$^1$& -31:07:35.7 & -09:29:02.2 & -38:08:27.2\\
		Vmag & 9.33(02)$^2$ & 10.865(38)$^3$ & 11.20(12)$^2$\\
		Distance [pc]$^4$ & 147.29(84) & 206.0(1.8) & 257.7(2.2)\\
		$M_{\rm A}$ [$\rm M_\odot$]$^1$& 1.26(10) & 1.18(10) & 0.980(80)\\
		\hline
		{\it Binary parameters} & & &\\
        $P_{\rm bin}$ [day] & 12.6427937(17) & 26.338279(14) & 23.322972(22)\\
		$K_{\rm A}$ [$\rm{km\,s^{-1}}$] & 27.87218(52) & 23.17988(87) & 17.26333(52)\\
		$e_{\rm bin}$ & 0.308724(19) & 0.120452(38) & 0.073151(38)\\
		$\omega_{\rm bin}$ [rad] & 3.243410(80) & 1.09802(37) & 2.39117(75)\\
		$T_{\rm peri}$ [BJD$-2,450,000$] & 7934.64483(13) & 7839.4468(15) & 8258.5885(28)\\
		\\
		{\it System parameters} & & &\\
		$\gamma$ [$\rm{km\,s^{-1}}$] & 29116.0(10) & -55395.95(65) & 16558.05(62)\\
		$\sigma_{\rm jit}$ [$\rm m\,s^{-1}$] & 2.77$^{+0.45}_{-0.41}$ & 1.48$^{+1.32}_{-1.45}$ & 0.089$^{+1.153}_{-0.085}$\\
		\hline
		{\it Derived parameters} & & &\\
		$M_{\rm B}$ [$\rm M_\odot$] & 0.408(20) & 0.444(23) & 0.268(14)\\
		$a_{\rm bin}$ [$\rm AU$] & 0.1260(31) & 0.2037(53) & 0.1720(44)\\
		\hline
	\multicolumn{4}{l}{	{\bf Notes:} 1 - \cite{Martin2019}, 2 - \cite{Tycho-22000}, 3 - \cite{Munari2014}, 4 - \cite{GaiaDR22018}}
	\end{tabular}

\end{table*}

\section{RV analysis with \kima}\label{sec:kima}

For the analysis of the radial velocities and the calculation of detection limits, we use the \kima package presented in \cite{Faria2018}. The code models the RV timeseries with a sum of Keplerian functions from $N_{\rm p}$ orbiting planets, estimating the posterior distributions for all the orbital parameters. 

To sample from the joint posterior distribution, \kima uses the 
Diffusive Nested Sampling (DNS) algorithm from \cite{Brewer2011}. Together with posterior samples, DNS provides an estimate for the marginal likelihood, or evidence, of the model, which can be used for model comparison \citep[e.g.][]{Brewer2014, Feroz2011}. Fixing the values of $N_{\rm p}$ in sequence, we can use the ratio of the evidences to compare models with different number of planets. In addition, since DNS can be used in a trans-dimensional setting \citep{Brewer2015}, the number of planets in the system $N_{\rm p}$ itself can be a free parameter in the analysis, and its posterior distribution can be estimated together with that of the orbital parameters. The posterior probability for each $N_{\rm p}$ value then allows for the same model comparison, with the advantage of being obtained from a single run of the algorithm.

The DNS algorithm samples from a mixture of distributions which is not directly the posterior. A  total number of samples $N_{\rm s}$ from this target distribution results in a smaller number of effective samples $N_{\rm eff}$ from the posterior distribution. We obtain $N_{\rm eff}>20,000$ effective samples for each model, which is more than enough to accurately characterize the posterior.

Keplerian parameters are estimated from the effective posterior samples via the clustering algorithm \texttt{HDBSCAN} \citep{McInnes2017}. First, crossing orbits are removed from the posterior samples. Those are proposed Keplerian orbits that cross with each other, or those with an eccentricity which would cause their orbit to cross into the instability region of the binary and are therefore unphysical \citep[e.g.][]{Dvorak1989,HOLMAN1999,Doolin&Blundell2011, Mardling2013}. \texttt{HDBSCAN} is applied on the remaining posterior samples, highlighting dense regions in parameter space. The cluster corresponding to the Keplerian signal is plotted with the \texttt{Corner} package \citep{corner}. Parameters are determined as the $50^{\rm th}$ percentile of the cluster, with $1\sigma$ uncertainties estimated from the $14^{\rm th}$ and $84^{\rm th}$ percentiles.

To decide between competing models that fit our data we use the \textit{Bayes factor}. The Bayes factor (BF, here onwards) is the ratio of the Bayesian evidence of the two competing models, and provides a measure of the support in favour of one over the other \citep{RobertE.Kass;AdrianE.Raftery1995, Trotta2008}. Table \ref{tab:Jeffreys scale} adapted from \cite{Trotta2008} shows how the BF is measured and introduces the Jeffreys' scale \citep{Jeffreys1961} as a measure of evidence strength.

\begin{table}
	\centering
	\caption{Table of Bayes Factors (BF) along with corresponding probability, sigma values (standard deviations away from the mean of a normal distribution), and the ``Jeffreys' scale'' adapted from \citet{Trotta2008}.}
	\begin{tabular}{l l l l}
		\hline
		\hline
	    BF & Probability & sigma & Evidence strength\\
		\hline
		$\lesssim3$ & $<0.750$ & $\lesssim2.1$ & Inconclusive\\
		$3$ & $0.750$ & $2.1$ & Weak\\
		$12$ & $0.923$ & $2.7$ & Moderate\\
		$150$ & $0.993$ & $3.6$ & Strong\\
		\hline
	\end{tabular}
	\label{tab:Jeffreys scale}
\end{table}

To guide our identification of planetary candidates we follow the same Jeffreys' scale, but use custom thresholds that affect what response we have to the system. Like \cite{Trotta2008}, we use $\rm BF = 12$ as a threshold for 'moderate' evidence. To identify planetary candidates worthy of additional follow-up (to apply for additional telescope time for instance), we use $\rm BF > 6$. Regularly, $3\sigma$ marks a detection in Astronomy. To make sure we are over that value, we place a threshold at $\rm BF = 35$. Any system reaching that level continues to be observed as any other, but we analyse its data more regularly than once a year. To visually track increasing evidence for a planetary signal, we place an additional division at $\rm BF = 70$. Finally we place our upper threshold $\rm BF = 140$, after which observations cease to be blind and we start to target specific epochs that represent poorly sampled orbital phases, or alternate solutions (eccentricity, $2\times P_{\rm p}$, etc). We use the thresholds described here in our insertion/recovery exercise below.

Contrary to a regular planetary system, a circumbinary SB1 system is dominated by the reflex motion caused by the secondary star (typically tens of $\rm km\,s^{-1}$) and extremely well constrained by the data. In addition, we know there cannot be a planet at an orbital period $\lesssim 4~P_{\rm bin}$ due to the instability region \citep{Dvorak1989, HOLMAN1999, Doolin&Blundell2011, Mardling2013}\footnote{In practice, the stability limit is dependent on other parameters such as the eccentricities, orbital alignments, relative orbital phases and mean motion resonance, as most recently explored by \citet{Quarles2018}. It is therefore possible that \kima would return unstable circumbinary configurations. However, any such solutions would have their stability rigorously tested with N-body models prior to claiming a detection.}. In contrast, we have no idea whether a planet is within the system, or what its parameters might be. By default \kima applies a common prior for all the orbiting objects (secondary, or planet) in a given model, which is not well adapted to our case. Particularly, there is no doubt that there is a secondary star in the system, and as such no need for the nested sampler to test that hypothesis.

In this context, \kima can instead use the so-called \emph{known object} mode, where a specific set of priors can be placed on the binary properties, and another set of priors is used to explore the presence/parameters of putative planets. At each step \kima then fits for the binary (the \emph{known object}) and any additional Keplerian signal present in the data. Using this mode, the nested sampler only computes evidence for $\geq1$ Keplerian function, therefore only testing for the presence of circumbinary planets ($\geq0$ planets).

A typical \kima run with $200,000$ saves, resulting in $N_{\rm eff} \approx 25,000-35,000$ posterior samples (depending on data-set) takes $\approx 180$ minutes on a standard laptop computer.

\section{Methods \& Simulations}\label{sec:Methods}

Here, we describe our methods, which are used similarly on all the systems we monitor in the survey. First, we give a simple parameterisation of the model. Then, we detail and justify the priors we use within \kima (see also Table~\ref{tab:prior distributions}). We then demonstrate using  \kima to produce robust and fully Bayesian detection limits. 
Following that, we perform a more traditional insertion-recovery test of static Keplerian signals to compare with our \kima detection limits. We then test the validity of injecting static Keplerian signals in comparison to N-body simulations. Finally, we discuss how our process compares with previous methods for detection limits, both for single and binary stars.

\subsection{Model setup}\label{sec:model_setup}

With \kima we assume independent, static Keplerian orbits for the binary and any circumbinary planets. This neglects the fact that the orbits of the planet and binary will evolve through three-body interactions (e.g. \citealt{Martin&Triaud2014,Kostov2014}). \citet{Martin2019} however found such interactions to be negligible with respect to the CORALIE BEBOP survey. We will briefly verify if this assumption still holds with our more precise HARPS data in Sect.~\ref{subsec:inj+rec}, where we test the injection and retrieval of N-body simulated RV signals.

The Keplerian models we fit to the data are defined by the following parameters. For the binary we have $P_{\rm bin}$ (the orbital period), $K_{\rm A}$ (the semi-amplitude of the primary star, caused by the secondary), $e_{\rm bin}$ (the binary's orbital eccentricity), $\omega_{\rm bin}$ (the argument of periastron) and $\phi_{\rm bin}$ (the starting phase of the orbit). We can calculate the time of periastron passage from these parameters using ${T_{\rm peri}} = t_0 - (P\times\phi)/(2\pi)$, where $t_0$ is the chosen epoch. 
The number of planets in the system is referred to as $N_{\rm p}$. Planetary parameters are defined like for the binary, but written with a subscript p, e.g. $P_{\rm p}$ for a planet's orbital period. 

In addition, we fit for the systemic velocity $\gamma$, and for a {\it jitter} term, $\sigma_{\rm jit}$ that rescales the uncertainties on the data. An increase in this parameter is penalised when computing the likelihood.
We can also fit for offsets in data between instruments, but this not necessary for the three systems we analyse here.

\subsection{Prior distributions}\label{sec:priors}
We use priors similar to those laid out in \cite{Faria2020}, but adapted to circumbinary planets. As described in Sect.~\ref{sec:kima}, we treat the signal produced by the secondary star as a {\it known object}, taken to be obviously present in the data and place tight priors on its parameters. We only compute Bayesian evidence for any additional signals to the inner binary. This is an advantage of using \kima, both the orbit of the binary and any additional signal are fit to the data simultaneously, each posterior sample obtained has a corresponding binary fit. Table~\ref{tab:prior distributions} shows the prior distributions utilised in our analysis. 

The secondary's orbital parameters prior distributions are based on values from an initial fit. We take the mode of each binary parameter and set to the prior to uniformly explore a range around that value. That range is determined from fitting all of the binaries in our sample. For instance, we explore $\pm0.01~\rm day$ around the orbital period, and  $\pm10~\rm m\,s^{-1}$ around the semi-amplitude (Table~\ref{tab:prior distributions}). While these might seem tight, the typical precision obtained on $P_{\rm bin}$ and $K_{\rm A}$ is $100\times$ less than the prior ranges we set (see Table~\ref{tab:System_parameters}). As such the binary priors we chose are wide enough to enclose the true parameters and their uncertainties.

To search for additional signals besides the binary (i.e. circumbinary planets), we use a uniform distribution for $\gamma$, $\phi$, $\omega$, and $N_{\rm p}$, as there is no reason to favour any particular value within these parameter spaces.

For planetary eccentricities, we utilise a Kumaraswamy distribution \citep{Kumaraswamy1980}, using values for our shape parameters as $\alpha=0.867$ and $\beta=3.03$ as justified in \cite{Kipping2013}, closely resembling the Beta distribution described there. A Kumaraswamy distribution favours lower values but still permits the exploration of higher eccentricities when the data allows.

The most sensitive parameter to sample when determining robust detection limits is planetary semi-amplitudes $K_{\rm p}$. For this reason, we use a transformed Log-Uniform (Jeffreys) prior. We cap the upper range, choosing $100~\rm m\,s^{-1}$ (any signal greater than this would be obvious in our data). Thanks to various tests, we noticed that setting a specific lower limit to $K_{\rm p}$ influenced our ability to recover planets and estimate robust detection limits. When we set the lower limit too high (e.g. $1~\rm{m\,s^{-1}}$) it excludes any signals $<1~\rm{m\,s^{-1}}$ in strength that could lie formally undetected in our data, but effectively ignoring any contribution they may have on other signals and parameters. If instead we set the limit too low (e.g. $0.01~\rm{m\,s^{-1}}$), the sampler does not explore the higher $K_{\rm p}$ sufficiently (since they become relatively less likely). This can bias the detection limit to lower $K_{\rm p}$ due to low number statistics in the posterior, producing over-confident detection limits.

Instead of setting a lower value we insert a {\it knee}, to give less priority to signals below a set value. \cite{Gregory2005} advises using a knee at $1~\rm{m\,s^{-1}}$, which was $1/10$ of their typical RV uncertainty. As our survey utilises RV data from both SOPHIE, HARPS and ESPRESSO, our uncertainties approach $1~\rm{m\,s^{-1}}$, we set our knee to $0.1~\rm{m\,s^{-1}}$. For values of $0~<~K_{\rm p} < 0.1~\rm{m\,s^{-1}}$ the distribution acts as a uniform prior, whereas for $K_{\rm p} \geq 0.1~\rm{m\,s^{-1}}$ the priors follow the usual Jeffreys prior \citep[see][]{Gregory2005}.

For all other parameters we employ Log-Uniform (Jeffreys) priors since the range of values can span several orders of magnitudes. This goes for $\sigma_{\rm jit}$ and $P_{\rm p}$.

The lower limit for the $P_{\rm p}$ prior is set to $4\times P_{\rm bin}$, following the instability region (see Sect.~\ref{sec:kima}). This is a slightly conservative estimate for the instability region. For completeness, we explored the effect of placing a lower limit at $P_{\rm p}< P_{\rm bin}$. Results were similar, with the only effect being that posterior samples were dispersed over a larger parameter space resulting in a coarser posterior.

We place the upper limit at $P_{\rm p} = \num{1e3}~\rm{day}$, roughly the time-span of our longest observed systems: J0310-31. For simplicity and to allow for a fair comparison, this limit is used for all three systems. We increase the maximum limit to $\num{2e4}~\rm{day}$ to generate detection limits, as described in Sect.~\ref{subsec:Detlim}, and witness how the detectable semi-amplitude increases with increasing orbital period after exceeding the time-span of the data. When seeking planetary candidates, we prefer using a maximum period of $\num{1e3}~\rm{day}$, which saves computational power and ensures a finer sampling of the posterior. The presence of a planetary signal with $P_{\rm p} > \num{1e3}~\rm{day}$ in our data would be identifiable as an over-density of posterior samples at the upper limit of this prior. In which case we adjust this limit as required. The limit on $P_{\rm p}$ will extended as we accumulate more data.

\begin{table}
	\centering
	\caption{Prior distributions used in RV model for binary and planetary signals in \kima}
	\begin{tabular}{c c c c}
		\hline
		\hline
	    Parameter & Unit & \multicolumn{2}{c}{Prior distribution}\\
		\hline
		    &   & Binary & Planet\\
		$N_{\rm p}$ &  & $1$ & $\pazocal{U}(0,3)$ or $1$\\
		$P$ & days & $\pazocal{U}(P_{\rm bin}\pm{0.01})$ & $\pazocal{LU}(4\times P_{\rm bin},\num{1e3}$ or $\num{e4})$\\
		$K$ & $\rm{m\,s^{-1}}$ & $\pazocal{U}(K_{\rm A}\pm{10})$ & $\pazocal{MLU}(0.1, 100)$\\
		$e$ & & $\pazocal{U}(e_{\rm bin}\pm{0.0005})$ & $\pazocal{K}(0.867, 3.03)$\\
		$\phi$ & & \multicolumn{2}{c}{$\pazocal{U}(0,2\pi)$}\\
		$\omega$ & & \multicolumn{2}{c}{$\pazocal{U}(0,2\pi)$}\\
		$\sigma_{\rm jit}$ & $\rm{m\,s^{-1}}$ & \multicolumn{2}{c}{$\pazocal{LU}(0.001,10\times \rm{rms})$}\\
		$\gamma$ & $\rm{m\,s^{-1}}$ & \multicolumn{2}{c}{$\pazocal{U}(V_{sys}\pm{100})$}\\
		\hline
	\end{tabular}
	\label{tab:prior distributions}
    \begin{tablenotes}
    \small
    \item \textbf{Notes:} $N_{\rm p}$ denotes the Number of Planetary Keplerian signals to fit to the data. $P_{\rm bin}$ and $e_{\rm bin}$ denote the Period and eccentricity of the binary respectively. $K_{\rm A}$ denotes the semi-amplitude of the primary star, caused by the secondary. $\pazocal{U}$ denotes a uniform prior with an upper and lower limit, $\pazocal{LU}$ is a log-uniform (Jeffreys) prior with upper and lower limits, $\pazocal{MLU}$ is a modified log-uniform prior with a knee and upper limit, and $\pazocal{K}$ is a Kumaraswamy prior \citep{Kumaraswamy1980} which takes two shape parameters.
    \end{tablenotes}
\end{table}

\subsection{Detection limit method}\label{subsec:Detlim}

A unique feature of a diffusive nested sampler such as \kima is that when forced to explore higher $N_{\rm p}$ than is formally detected, the sampler will produce a map of all signals that are compatible with the data. Since those proposed signals remain formally undetected, the posterior naturally produces a detection threshold. In our case this is made simpler by \kima's {\it known object} mode (see Sect.~\ref{sec:kima}), and the fact that no planets have been detected in those three systems thus far. We describe in Sect.~\ref{sec:rebound_recovery} how we calculate a detection limit when a planetary signal is already present in the data.

Later, we compare our results to a more traditional insertion-recovery test, the method of choice to assess detectability and measure occurrence rates in radial-velocities surveys \citep[][]{Cumming2008, Konacki2009b, Bonfils2013, Martin2019, Rosenthal2021, Sabotta2021}. 

Using a diffusive nested sampler like \kima also provides immediate advantages over insertion-recovery test. \kima samples all orbital parameters in a fine manner. Particularly, this means that all orbital phases, all eccentricities and all periastra are sampled when typically only a small number of phases are tested, and that eccentricities are nearly always forced to zero. Detection limits obtained with \kima marginalise over more orbital parameters than is computationally feasible with insertion-recovery tests, and \kima assumes Keplerian signals, rather than sinusoidals (see next section).

To create a detection limit, we run \kima with priors as in Table~\ref{tab:prior distributions} except that we fix $N_{\rm p} = 1$. \kima marginalises over all allowed parameter space, searching for any possible Keplerian signals producing a reasonable likelihood, in the process ruling out Keplerians that would otherwise have been detected. 

\begin{figure*}
	\includegraphics[width=\textwidth]{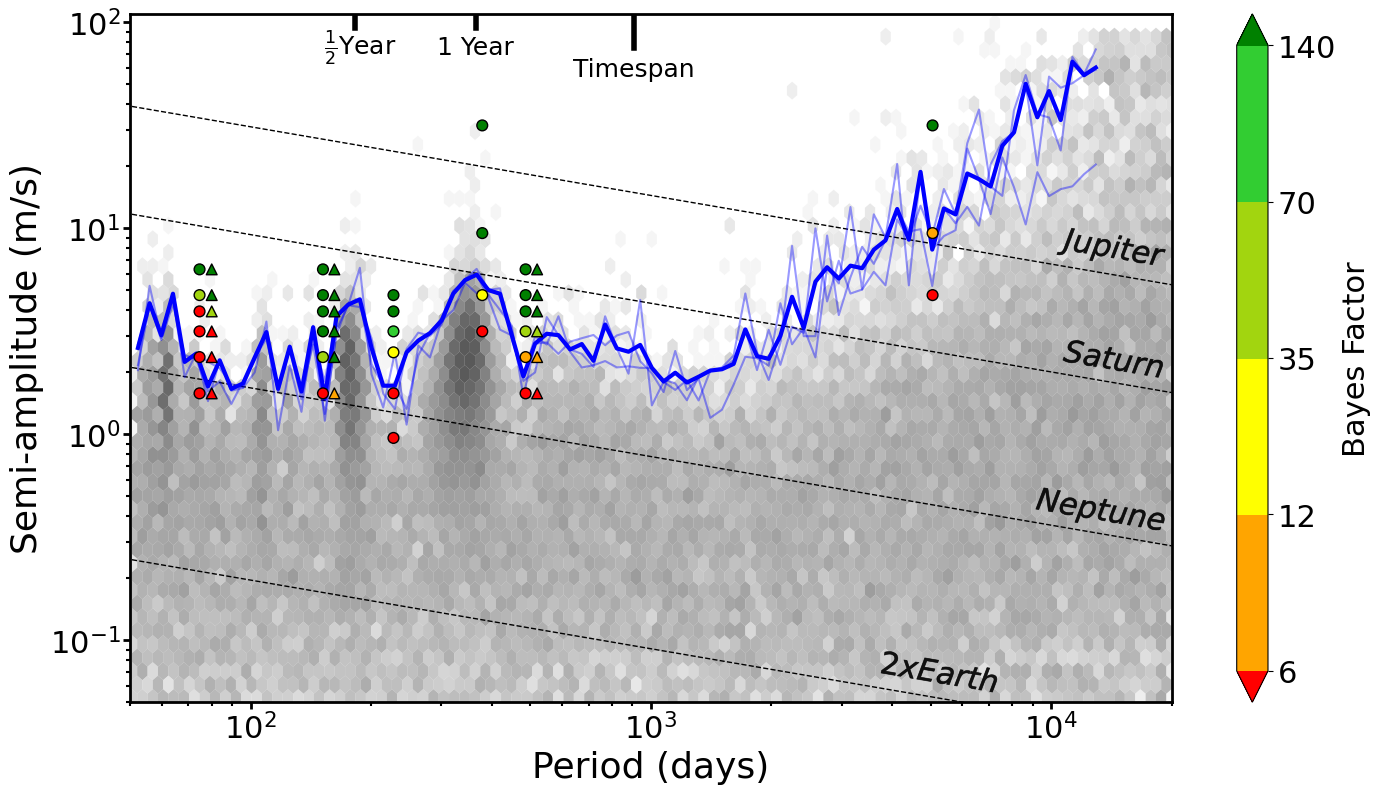}
    \caption{Hexbin plot denoting the density of posterior samples obtained from three separate \kima runs on J0310-31 with $N_{\rm p}$ fixed $=1$, as a grayscale. Faded blue lines show detection limits calculated from each individual run on the system as described in the text. The solid blue line shows the detection limit calculated from all posterior samples combined. Coloured symbols indicate the Bayes factor results of injection/recovery tests and correspond to the colour bar on the right. Circles are for injected static Keplerian sinusoidal signals, whereas triangles are for N-body simulated injected signals. The Keplerian and N-body signals are injected at the same orbital periods but are represented slightly offset horizontally here for visual clarity. The faded red dashed lines show masses of solar system planets for comparison.}
    \label{fig:J0310-31_Det_lims}
\end{figure*}

A key to producing a robust detection limit is to produce a well sampled posterior. For each system we compute three runs, ensuring $>20,000$ samples are obtained in each. To gather consistent results we set the number of saves in \kima to $200,000$. The resulting ($K_{\rm p}, P_{\rm p}$) posterior's density is plotted in Figures~\ref{fig:J0310-31_Det_lims} and \ref{fig:J1540_andJ1920_Det_lims} as a greyscale hexbin, with the detection limit being the top envelope.

To compute the detection limit, we separate the posterior in log-spaced bins in $P_{\rm p}$, and within each bin we evaluate the maximum $99^{\rm th}$ percentile of the $K_{\rm p}$ distribution.
This is done with the caveat that $\geq 2$ posterior samples must lie above the chosen sample in a given bin. We do this to prevent the detection limit from being affected by small number statistics in individual bins.
We calculate a detection limit for each of the three runs we perform on each system, which we show with faded blue lines in Figures~\ref{fig:J0310-31_Det_lims} and \ref{fig:J1540_andJ1920_Det_lims}. We also draw the detection limit obtained when combining all posterior samples from the three runs into a single sample, with a solid blue line. Proceeding this way allows us to produce a mean detection limit and to obtain a visual estimate of the uncertainty of that limit for each bin. Overall, all runs are compatible with one another. We stop computing the detection limit for all $P_{\rm p}$ exceeding the first $P_{\rm p}$ bin where the number of samples in the top $10\%$ of $K_{\rm p}$ prior is larger than three, which is where the posterior is affected by the upper limit set for the $P_{\rm p}$ prior.

\subsection{Injection and Recovery Tests}\label{subsec:inj+rec}

The seminal (single-star) radial-velocity surveys of \citet[][]{Cumming2008, Mayor2011, Howard2010} used injection and recovery methods as a means of determining injection limits. First, all known planets are removed from the data. Then a sinusoid of varying amplitude, period and phase is used to model a putative exoplanet, and applied to the data. Following, that, a periodogram of the data (typically a Generalised Lomb-Scargle) is computed. If it produces a peak below a certain false alarm probability (typically $<1\%$) near the injected period, the simulated planet is considered ``detected''. By reproducing this procedure over a grid of $P_{\rm p}$ and $K_{\rm p}$ inserted signal, it is possible find out for which values the planet is no-longer detectable. \citet{Martin2019} followed this approach for the CORALIE BEBOP survey, with one key difference that the planet signal was injected to a radial velocity data set where the binary signal had already been removed.

Whilst this traditional approach may appear similar to ours, in practice they are distinct. When we use our Bayesian approach with \kima to produce detection limits we answer the question "what is compatible with the data?", whereas with an insertion-recovery test, we ascertain "can this \textit{specific} signal be found?".

Here we create injection-retrieval tests for our three targets. There are two purposes of this. First, we investigate compatibility between our Bayesian \kima-derived limit and the more traditional approach of injecting static Keplerian signals. Second, we follow the approach of \citet{Martin2019} to inject not just static sinusoids but an N-body derived signal using \rebound, including all dynamical interactions between the planet and binary, to test the validity of the previous step.

We add one circumbinary planet signal to the original data, and do so for a number of planetary semi-amplitudes $K_{\rm p}$ and orbital periods $P_{\rm p}$. \kima is then used to recover any Keplerian signals in the modified data-set with the resulting Bayes factors providing a measure of signal recovery success.

The choice of $K_{\rm p}$ and $P_{\rm p}$ at which we inject simulated signals is informed by the data itself. For $K_{\rm p}$, values are chosen as multiples of the residual RMS of each system. Values used can be found for each system in 
Tables~
\ref{tab:0310_injrec}-\ref{tab:1928_injrec}. 

For $P_{\rm p}$ there are three periods we test for all three targets. The first is $5.9\times P_{\rm bin}$, which is similar to known circumbinary systems \citep{Martin2018} and slightly offset from the often unstable 6:1 mean motion resonance \citep{Quarles2018}. We also test $12\times P_{\rm bin}$ and $486~\rm{days}$ ($365\times4/3$, to avoid but be close to the $1~\rm yr$ alias). In some cases, we also simulate specific orbital periods, coinciding with features of the detection limits described in Sect.~\ref{subsec:Detlim}. 

Each inserted signal has an eccentricity $e_{\rm p} = 0.01$ (small but non-zero), and argument of periastron $\omega_{\rm p} = 0$.

\begin{figure*}
	\includegraphics[width=\textwidth]{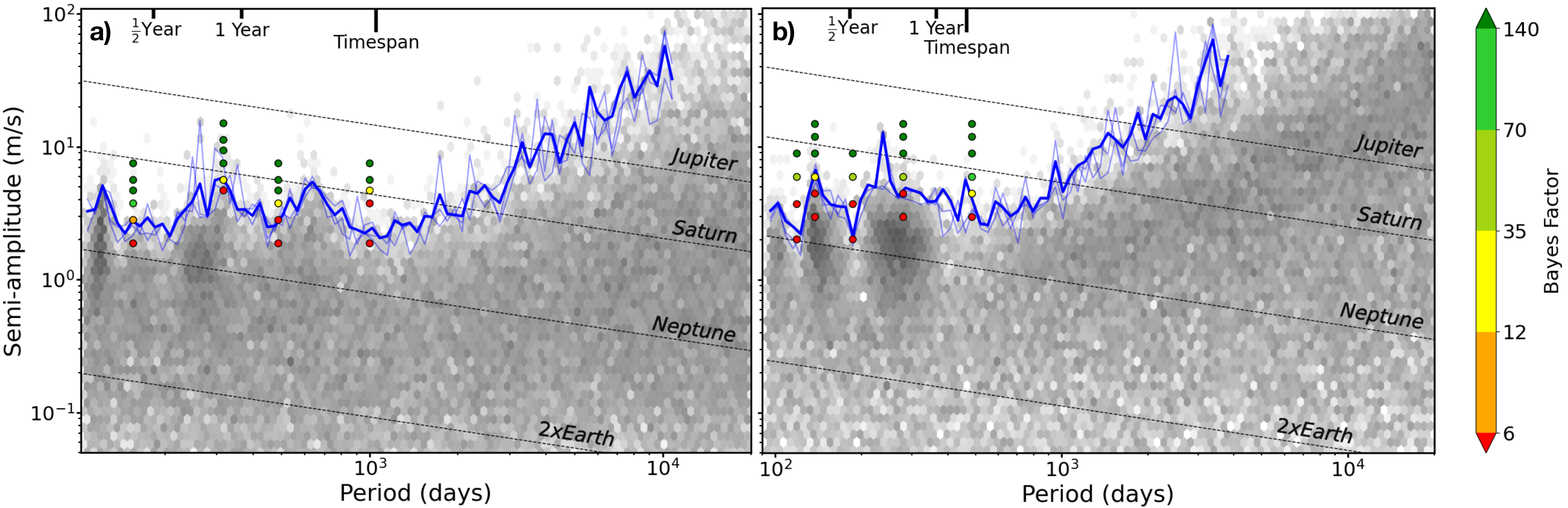}
    \caption{Same as in figure~\ref{fig:J0310-31_Det_lims} but for three separate \kima runs on J1540-09 (a), and J1928-38 (b) with $N_{\rm p}$ fixed $=1$. All injected signals here are static Keplerian sinusoidal.}
    \label{fig:J1540_andJ1920_Det_lims}
\end{figure*}

Then, we run \kima as described previously, using priors as in Table~\ref{tab:prior distributions}, including a uniform prior on $N_{\rm p} = U(0,3)$, reproducing how we conduct an open search of the system. Bayes factors are computed from each run and results are plotted in Figures~\ref{fig:J0310-31_Det_lims} and \ref{fig:J1540_andJ1920_Det_lims} as coloured points. Their values can also be found in Tables~
\ref{tab:0310_injrec} to \ref{tab:1928_injrec}.

The binary and putative circumbinary planets for the three chosen periods are then simulated for the J0310-31 system using the \rebound N-body code \citep{Rein&Liu2012}, with the IAS15 integrator \citep{Everhart1985, Rein2015}. First we remove the best fit binary model from J0310-31's observed radial-velocity measurements, producing residuals. Then we use \rebound to simulate a radial-velocity model of the binary and putative planet at each epoch of observation. We add this model to the residuals. Doing this preserves the scatter, uncertainties and observational cadence obtained in reality. \kima is ran once again  on each N-body simulated dataset, and Bayes factors computed and compared to those resulting from the injected static Keplerian signals.We discuss the result of this test in Sect.~\ref{sec:rebound_recovery}. All Bayes factor values are plotted in Fig. ~\ref{fig:J0310-31_Det_lims} as coloured triangles offset to the right, and can also be found in 
table~
\ref{tab:0310_injrec_rebound}.

\section{Results and Discussion}\label{sec:results_dis}

We begin by providing updated parameters on the three binary systems investigated. Then, we discuss the calculated detection limits for each binary system, along with how they compare to our Keplerian injection and recovery tests. Following this we discuss and present our results from the comparison between Keplerian and N-body simulation injection and recovery. We discuss how nested sampling provides more robust detection limits. Particularly, we describe how  traditional insertion-recovery exercises over-estimate their ability to retrieve planets by assuming zero eccentricity.

\subsection{Updated Binary Parameters}\label{sec:bin_params}

Table~\ref{tab:System_parameters} shows updated orbital parameters from our most recent data for the systems investigated in this work. We provide $1\sigma$ uncertainties for each parameter, in parenthesis. Uncertainties are symmetric and varying only in the second significant figure, where the largest of the two values were taken. $\sigma_{\rm jit}$ is the only parameter that demonstrated an asymmetric distribution (tending to zero), where we consequently provide asymmetric uncertainties.
Values agree with \cite{Martin2019} to within $1-2\sigma$ for all parameters other than $\omega_{\rm bin}$ and correspondingly $T_{\rm peri}$. Improvements in parameter precision of around $70\%$ are be attributed to the increased number of high-precision measurements acquired using the HARPS spectrograph.

\subsection{Detection limit results}\label{sec:det_lim_results}

Figures~\ref{fig:J0310-31_Det_lims} and \ref{fig:J1540_andJ1920_Det_lims} show the results of our detection limit analysis for EBLM~J0310-31, J1540-09, and J1928-39 respectively. The greyscale \textit{hexbins} show the density of posterior samples from all \kima runs on the targets. The faded blue lines are the calculated detection limit from three individual run containing $200,000$ saves, and $>20,000$ posterior samples. The solid blue line is calculated from all posterior samples combined. Coloured dots represent the results for each individual injection and recovery test as outlined in Sect.~\ref{subsec:inj+rec}.
The colour of these points represent the Bayes Factor (BF) discussed in Sect.~\ref{sec:kima}, and measure the probability of recovery for the injected Keplerian signals. Every sample in the posterior used to compute detection limits involves a free fit of the binary as a well as a proposed planetary signal, producing a detection limit that marginalises over the binary parameters, an essential element for BEBOP.

From these plots, it is evident that different runs of \kima produce consistent detection limits with one another. The only inconsistencies are due to a low number of samples within a particular bin, a situation easily resolved by increasing the number of saved posterior samples. We also note that the posterior below the limit is uniformly sampled across parameter space, an indication of the reliability of the method we followed.

The detection limit plots also show many features that are regularly seen in detection limits for exoplanets using the radial-velocity method. For all three systems we see an increased density of posterior samples, and a consequently raised detection limit near $0.5$ and $1~\rm yr$ orbital periods, corresponding to yearly aliases, caused by seasonal gaps in observations. In addition, we observe that our detection capability is broadly horizontal for $P_{\rm p}$ below the timespan of our data. Then, it increases linearly in log until it hits the upper limit of the $P_{\rm p}$ prior. For J0310-31, we reach a mean $K{\rm p}$ detection of $1.49~\rm m\,s^{-1}$, for J1540-09, we get $2.06~\rm m\,s^{-1}$, and $2.15~\rm m\,s^{-1}$ for J1928-38. For these three systems, we outperform any produced by the TATOOINE survey \citep{Konacki2010} despite our systems being fainter by an average four magnitudes, validating our choice to monitor single-lined binaries to seek circumbinary exoplanets. These detection limits can be translated into masses, which are shown by additional lines in Figures~\ref{fig:J0310-31_Det_lims} and \ref{fig:J1540_andJ1920_Det_lims}. Typically, we are sensitive to planets with masses between Neptune's and Saturn's for orbital periods between 50 to 1000 days respectively, a milestone.

Injection-recovery test results are consistent with the detection limits obtained with \kima at almost all orbital periods tested on each system. The largest deviation can be seen in Fig.~\ref{fig:J0310-31_Det_lims} for J0310-31 at $74.59~\rm{d}$. Where the detection limit is approximately $2~\rm{m\,s^{-1}}$ lower than the recovered injected signal strength. This is the only location where such a deviation is observed. 

\begin{figure}
	\includegraphics[width=\columnwidth]{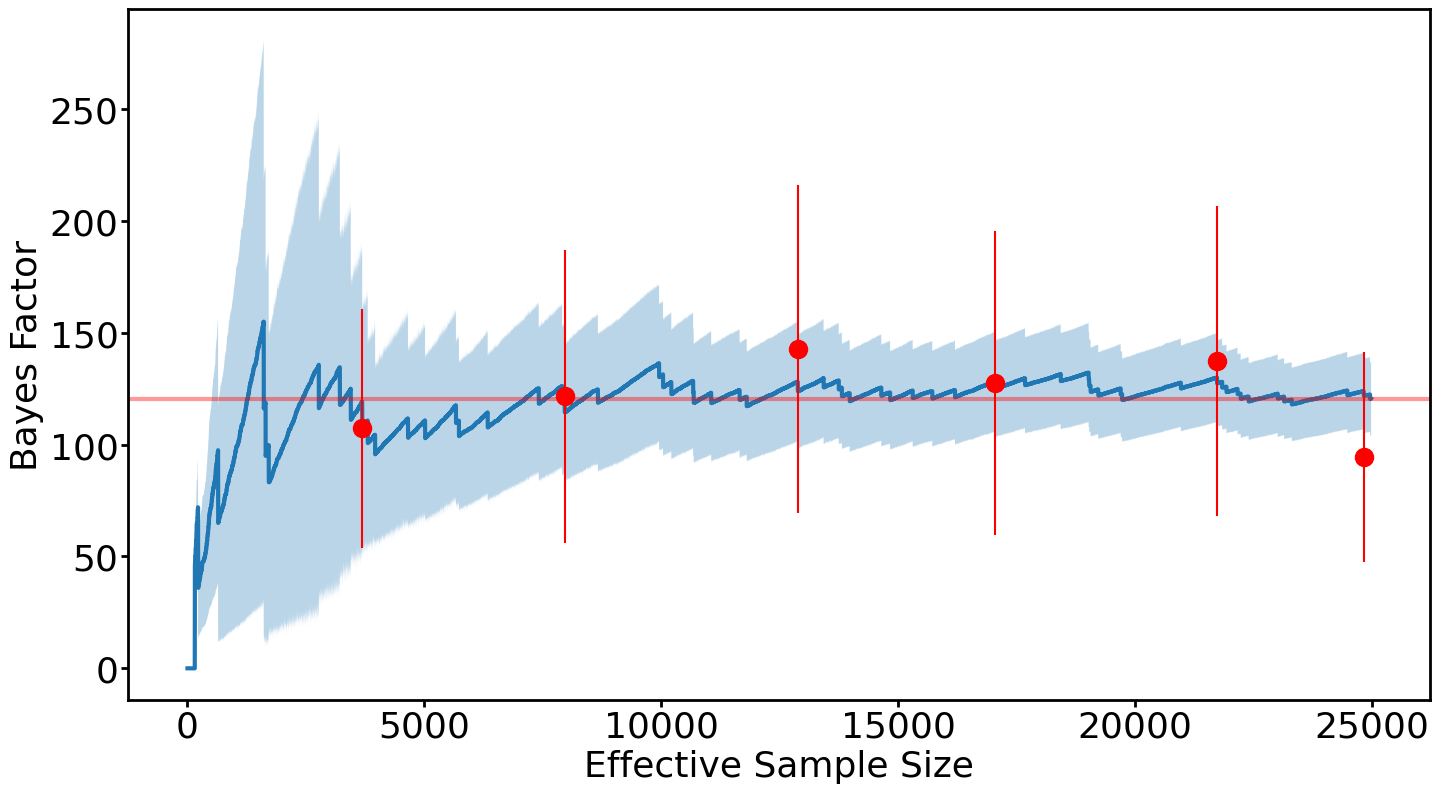}
    \caption{Plot of Bayes Factor (BF) vs number of posterior samples from a \kima run with a $227.57~\rm{d}$, $3.16~\rm{m\,s^{-1}}$ injected signal in J0310-31. In blue, we show the cumulative Bayes factor as a function of the number of posterior samples. The blue line shows the BF values, and and their running uncertainty are drawn as the shaded blue region. We add one red dot after $10$ $N_{\rm p}=0$ samples to show the BF value then, with their uncertainties. The horizontal red line shows the final BF value calculated with all posterior samples.}
    \label{fig:BFvsample}
\end{figure}

Figure \ref{fig:BFvsample} is an example of how the value of the Bayes Factor evolves with the number of posterior samples during a \kima run. Our example is J0310-31 data, with an injected planetary signal with $P_{\rm p} = 227.57~\rm{d}$, and  $K_{\rm p}= 3.16~\rm{m\,s^{-1}}$. The uncertainty on the Bayes Factor (blue area) is estimated assuming a multinomial distribution. We use this figure as an indicator for fit convergence, i.e once the line asymptotes on a BF value the fit has converged.

We also check how often we recover the Keplerian signal we insert. We find that orbital parameters of injected signals detected with $\rm BF > 35$ are typically recovered to within $3\sigma$ or $10\%$ of their original values, except those with periods close to one or half a year. This is true for injected Keplerian signals, as well as the {\sc rebound} injected signals.

\begin{figure}
	\includegraphics[width=\columnwidth]{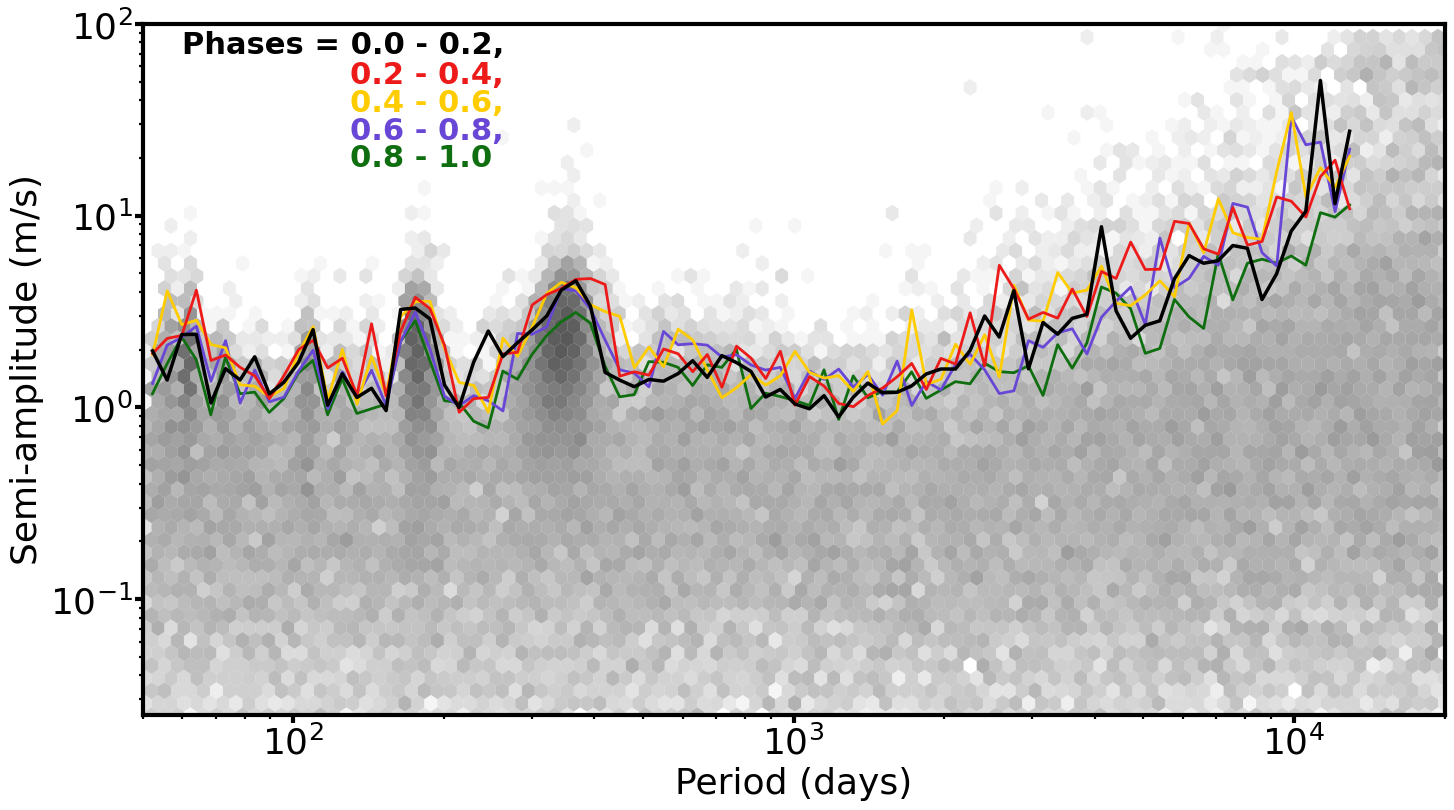}
    \caption{Hexbin plot of posterior samples for J0310-31 as in figure~\ref{fig:J0310-31_Det_lims}. The coloured lines here indicate the detection limits calculated with the posterior samples split into five even phase bins with the black line representing the phase bin containing $\phi = 0$.}
    \label{fig:J0310-31_Det_lims_phase}
\end{figure}

The chosen phase of the injected Keplerian signal could in principle cause the discrepancy between the methods seen at $P_{\rm p} = 74.59~\rm day$ for J0310-31. Figure~\ref{fig:J0310-31_Det_lims_phase} shows the same posterior samples as those in Fig.~\ref{fig:J0310-31_Det_lims} but split into five different phase bins, with a detection limit calculated for each. Doing this is simple with \kima, we simply select all posterior sample within a specific phase bin and reproduce the method in Sect~\ref{subsec:Detlim}. As can be seen in the figure, varying phase has little effect on the detection limit, which is the case for each target investigated.
This consistency seen between phase bins further demonstrates how robust and consistent our detection limits are. This also highlights the usefulness and importance of using nested samplers such as \kima to compute detectability curves.

\subsection{Rebound Simulation Results Comparison}\label{sec:rebound_recovery}
Results from systems simulated using \rebound are shown in Fig.~\ref{fig:J0310-31_Det_lims}, as coloured triangles offset to the right from similar, but Keplerian-only simulations.
The only evident deviation between the two simulations occurs at the shortest period investigated ($74.59~\rm{day}$) here the Bayes factor for the \rebound simulated data is higher for $K_{\rm p} = 3.95~\rm{m\,s^{-1}}$.

The agreement in those results (see Table~
\ref{tab:0310_injrec_rebound}) 
confirms that assuming Keplerian functions has no obvious detriment to circumbinary planet detection within our current data just like for any planetary system so far, and as we had already seen in \citet{Martin2019} with more imprecise CORALIE data. This simply means that assuming a static Keplerian signal is sufficient for discovery. However, dynamical fits are known to be useful to constrain the physical and orbital parameters of circumstellar planetary systems \citep[e.g. GJ~876;][]{Correia2010}, and would likely be the case for circumbinary systems, as they have been for HD~202206 \citep[][]{Correia2005, Couetdic2010}, a system comparable to a circumbinary configuration.

\begin{figure}

	\includegraphics[width=\columnwidth]{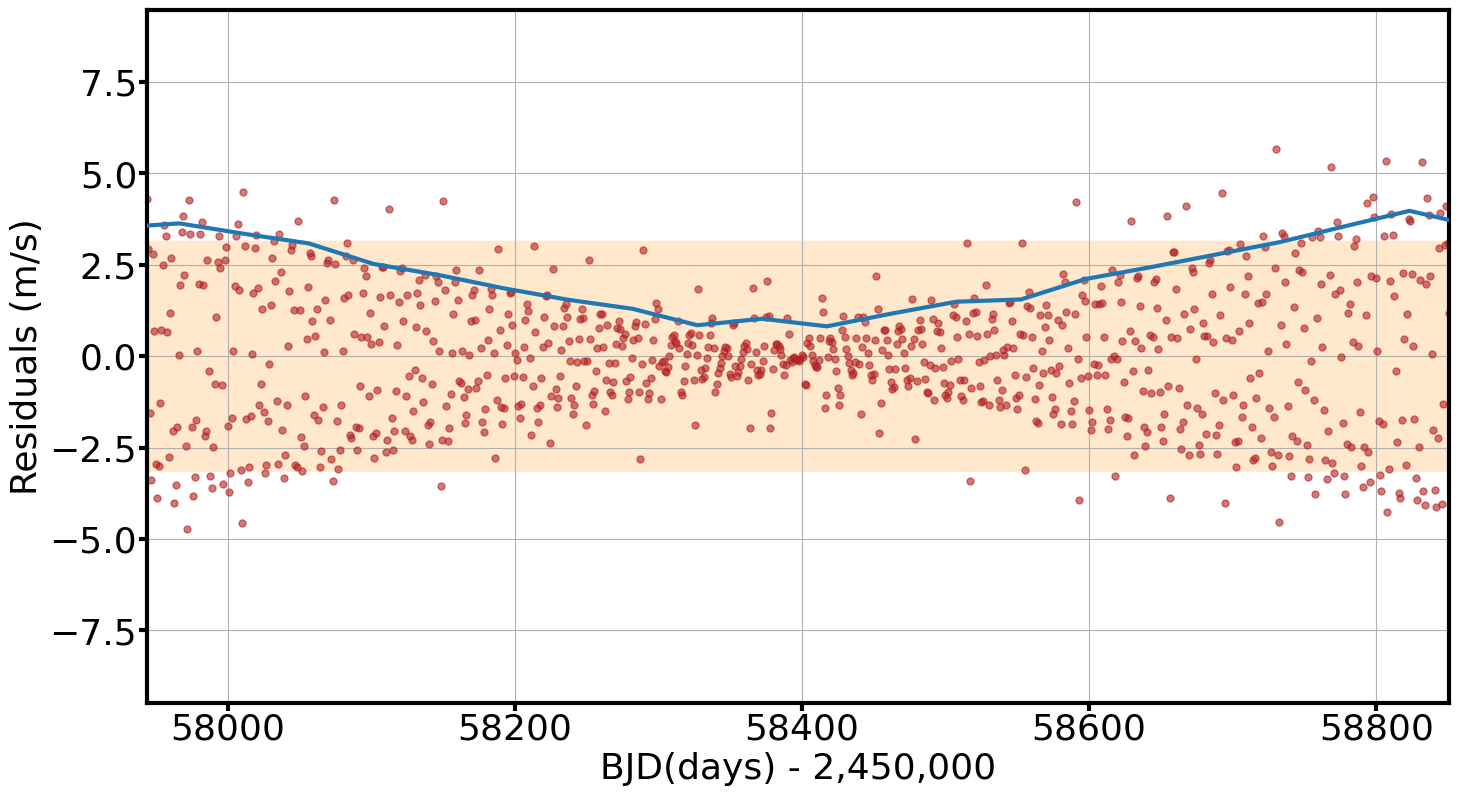}
    \caption{Plot of residual radial velocity amplitude as a function of time, after removing a Keplerian binary signal from N-body simulated J0310-31 radial-velocity dataset. Simulated data are computed with  \rebound including the binary and an orbiting circumbinary planet with $P_{\rm p} = 74.59~\rm{d}$ and $K_{\rm p} = 6.32~\rm{m\,s^{-1}}$. To reveal the Newtonian perturbation, we plot one point for each night over the entire duration of our dataset. The blue line depicts $1.3 \times$ RMS within bins $45~\rm day$ wide. We use this line as a visual guide to the amplitude of the effect. The shaded orange region illustrates the RMS scatter obtained on the  data we collected on J0310-31. The {\it bow-tie} shape of the residuals is caused by apsidal motion of the binary ($\dot{\omega}_{\rm bin}$), caused by the planet, which a Keplerian model does not include. }
    \label{fig:J0310-31_newt}
\end{figure}

To better understand the non-Keplerian signal, we visualise the amplitude of the Newtonian interactions in Fig.~\ref{fig:J0310-31_newt}. We use \rebound to simulate nightly RV data, for $K_{\rm p} = 6.32~\rm m\,s^{-1}$ at  $P_{\rm p} = 74.59~\rm days$ (the shortest period and highest mass planetary signal simulated in this work, producing the largest Newtonian perturbation), over the timespan of our data on J0310-31. Figure~\ref{fig:J0310-31_newt} depicts the residuals after fitting and removing the Keplerian binary and planetary signals from the N-body simulated dataset. The residuals show that N-body model diverges from the Keplerian model due to apsidal precession ($\dot{\omega}_{\rm bin}$). However, for the timespan of our current data that divergence remains comparable to the residuals' RMS. Assuming Newtonian effects are non-negligible will likely cease to be valid for longer time-series. 

\subsection{Post-Newtonian Effects}\label{sec:non-kep}

Radial velocity measurements of binary stars are also affected by weaker effects, such as tidal distortion, gravitational redshift, transverse Doppler and light time travel effects \citep{Zucker2007, Konacki2010, Arras2012, Sybilski2013}. \kima assumes purely Keplerian functions, ignoring these effects, here we explore what impact this may have on our fit, and particularly on our ability to retrieve planets.

We calculate the magnitude of tidal distortion on our RV measurements as in \cite{Arras2012}, and find their amplitude is $<0.7~\rm m\,s^{-1}$ for the three systems we investigate in this paper, an insignificant contribution to the observed RV scatter. We do note however, that for the shortest period binaries in the BEBOP survey, the magnitude of this effect becomes significant (with respect to the RMS of the systems) and should be accounted for in any further analysis.

Computing equations from \cite{Zucker2007} for the parameters of each of our three systems, we find the magnitude of transverse Doppler effects to be $<2.2~\rm m\,s^{-1}$, and the peak to peak variation $<1.6~\rm m\,s^{-1}$. Similarly, we find the magnitude and peak to peak variation of light time travel effects is $<2.7~\rm m\,s^{-1}$. The only relativistic effect found to be greater than the RMS on these systems is that of gravitational redshift, with a maximum amplitude of $13.9~\rm m\,s^{-1}$ but a peak to peak variation of $<6.5~\rm m\,s^{-1}$.

\begin{figure}

	\includegraphics[width=\columnwidth]{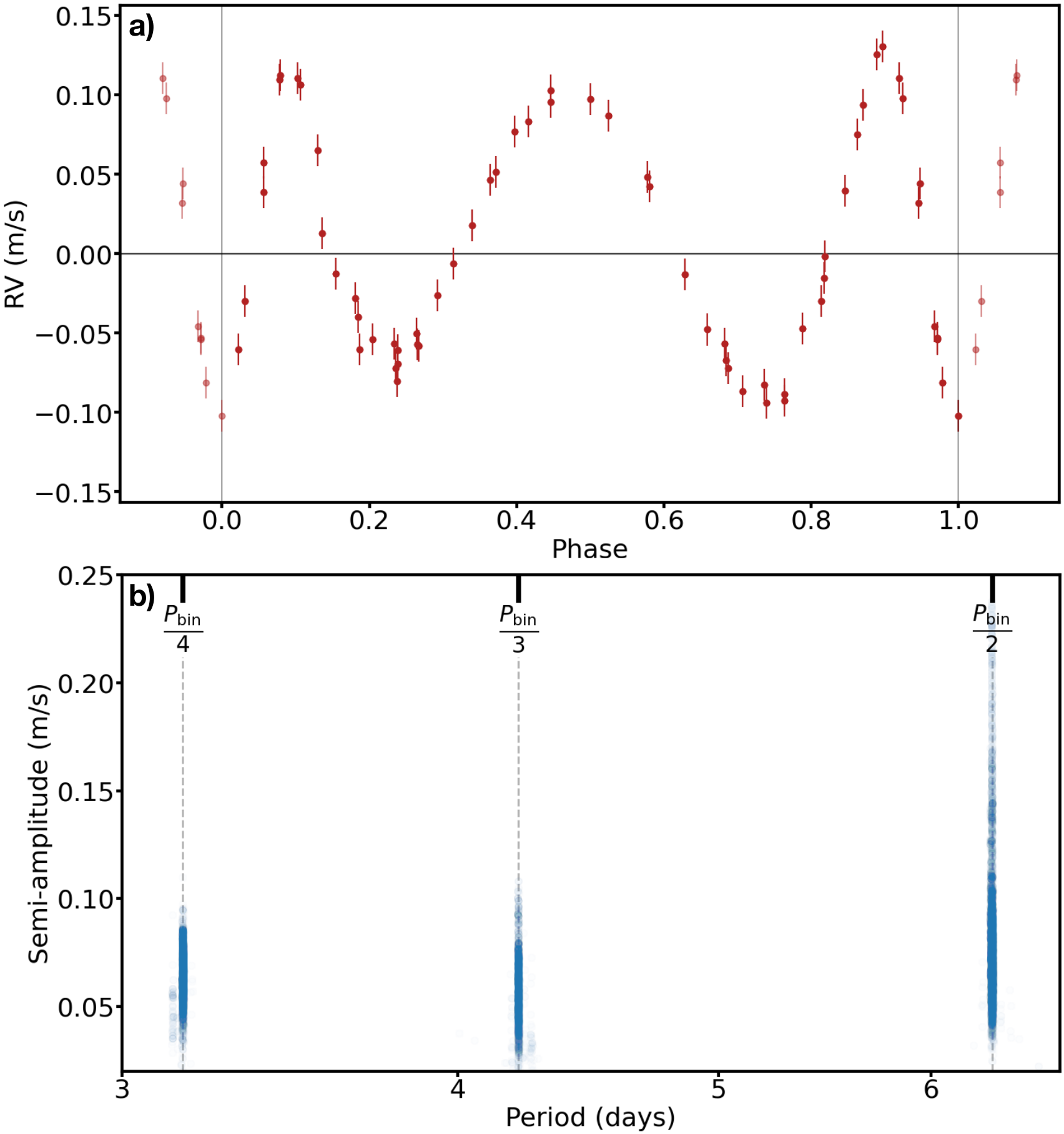}
    \caption{a) Phased RV residuals from the best fit model of J0310-31 simulated binary and additional post-Newtonian effects only. Residuals have an RMS $=~0.07~\rm m\,s^{-1}$.
    b) Semi-amplitude vs period plot of posterior samples from a \kima run on the residuals seen in plot a). Posterior samples are located at half, third and a quarter of J0310-31's binary orbital period.}
    \label{fig:J0310_non_kep_residuals}
\end{figure}

To better asses the impact of these effects on our work, we simulate a purely Keplerian binary signal for J0310-31 using the same cadence as our observations were obtained in, and add these post-Newtonian effects. We then fit the data using \kima. Figure~\ref{fig:J0310_non_kep_residuals}a shows residuals after removing the best fit Keplerian model of the binary. The residuals have an amplitude of $\approx0.1~\rm m\,s^{-1}$ and a signal phasing with the binary (as expected). This shows that most of the post-Newtonian effects are absorbed by the Keplerian fit.

We now use \kima once more, to "search for a planet" in these residuals, in order to study how the algorithm behaves in the presence of additional coherent signals. The posterior of that fit is plotted in figure~\ref{fig:J0310_non_kep_residuals}b, and demonstrates the periodicity of these remaining signals at harmonic periods of the binary with low semi-amplitude.

Individually these effects boast significant amplitudes, but as their functional forms resemble, and phase, with a Keplerian, only small differences are produced: Our simulations show that the majority of the post-Newtonian signals are absorbed into our fit. They are absorbed primarily into the systemic velocity, and lead to a small underestimation of the secondary's semi-amplitude. For J0310-31, ignoring post-Newtonian effects in this experiment creates an underestimated $K_{\rm A} \approx4~\rm m\,s^{-1}$, corresponding to a $0.01\%$ deviation.
We conclude that these effects do not affect the detection limits of this paper greatly, and can be safely ignored for our purposes.

\subsection{Detection limits in the presence of planetary signals}\label{sec:det_lim_revcovery}

\begin{figure}
	\includegraphics[width=\columnwidth]{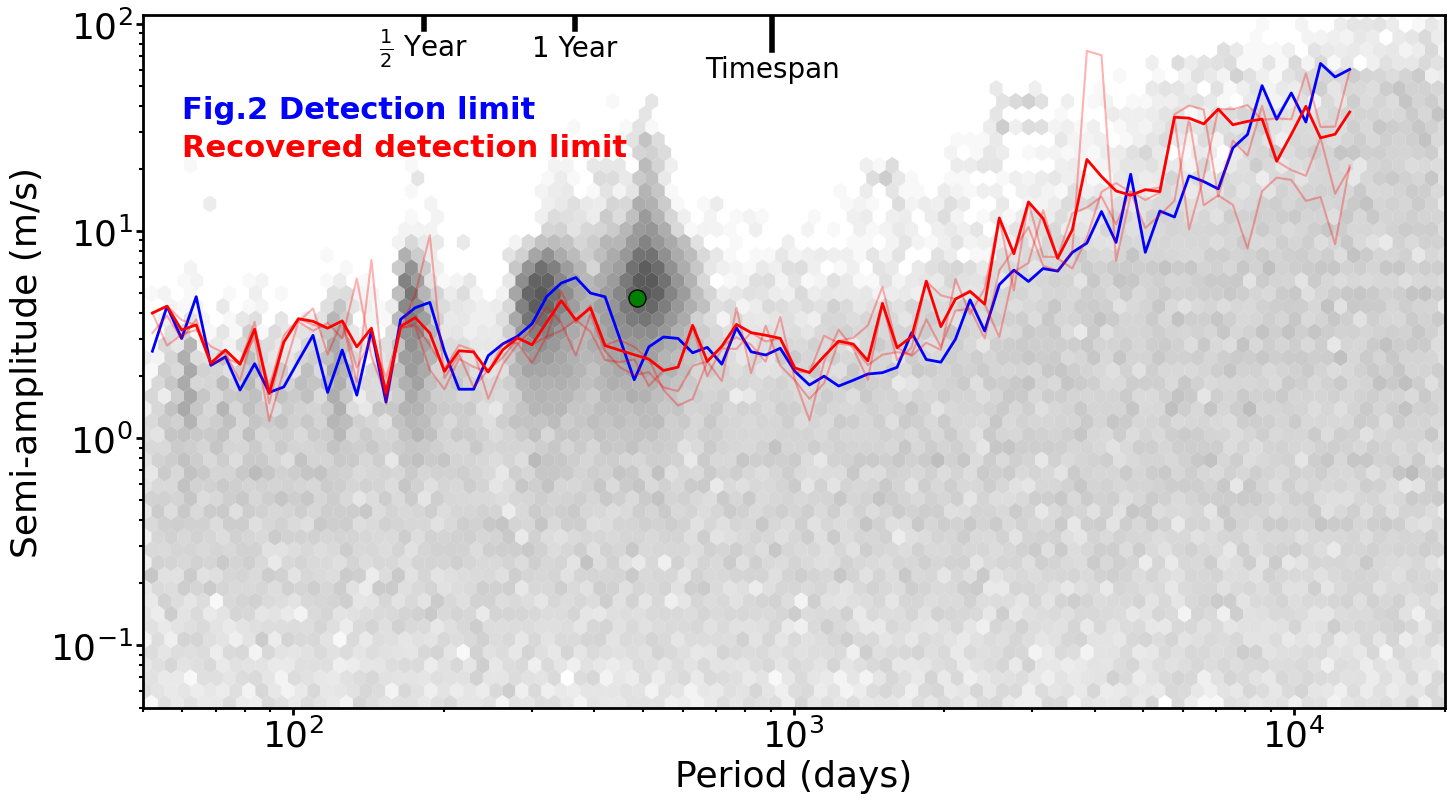}
    \caption{Hexbin plot of posterior samples obtained from three separate \kima runs on J0310-31 with an additional $486$d signal located at the green point, with $N_{\rm p}$ free between $0$ and $3$. The blue line shows the detection limit as seen in Fig.~\ref{fig:J0310-31_Det_lims}. Faded red lines show detection limits calculated from each individual run on the system after having removed the most likely $N=1$ signal from the data. The solid red combines all these posterior samples together.}
    \label{fig:J0310-31_recovered_det_lim}
\end{figure}

Detection limits thus far are all generated from our current data on systems purposefully chosen with no candidate planetary signals. The ultimate goal of the BEBOP survey is to detect RV signals of circumbinary planets in our data. Once a system is found to have a planetary signal, a detection limit is required to rule out the presence of additional companions above our calculated mass limit. 
To this end, we calculate the detection limit for J0310-31 with our N-body simulated data-set containing an injected $4.74 \rm m\,s^{-1}$, $486$~d signal. A signal of this strength is easily detected in the data (see Tab.~
\ref{tab:0310_injrec_rebound}
for its Bayes factor value).
We identify the injected signal as in Sect.~\ref{subsec:inj+rec}, with $N_{\rm p}$ varying uniformly from $0$ to $3$, as is standard for the survey. Figure~\ref{fig:J0310-31_recovered_det_lim} shows the resulting posterior samples, along with a green dot illustrating the semi-amplitude and orbital period of the injected signal. 
To obtain a recovered detection limit we subtract the Keplerian signal corresponding to the posterior model with the highest likelihood from the simulated data. Once subtracted, we run \kima again as in Sect.~\ref{subsec:Detlim} and calculate the red recovered detection limit from the resulting posterior samples.
The blue line seen in Fig.~\ref{fig:J0310-31_recovered_det_lim} is taken from Fig.~\ref{fig:J0310-31_Det_lims} for reference, and the recovered red line closely resembles the original blue detection limit.
We note that while this approach is not strictly Bayesian, it recovers the correct detection limit, and is expected to affect only a few systems (with planets) within the survey.

\subsection{The Dangers of Assuming Circular Orbits}
Thanks to our use of \kima to produce detection limits we are able to explore the effect of assuming $e_{\rm p} = 0$ on detection limits for exoplanet radial-velocity surveys.

Most often detection limits are produced with multiple insertion-recovery tests. To increase efficiency, and remain computationally tractable, a number of assumptions are made.
Circular orbits are generally assumed in the inserted signals used to calculate detection limits because this removes two dimensions: $e_{\rm p}$ and $\omega_{\rm p}$ \cite[e.g.][]{Cumming2008, Zechmeister2009, Howard2010, Bonfils2011, Mayor2011, Bonfils2013, Martin2019, Sabotta2021}\footnote{Circularity is assumed for recovery since most use a periodogram. With \kima the full Keplerian is used for exploration and recovery}. In addition, a short number of discrete planet phases $\phi_{\rm p}$ are usually sampled (e.g. 10; \citealt{Martin2019} or 12; \citealt{Zechmeister2009, Bonfils2011, Bonfils2013}). 

Commonly when orbits are assumed circular, investigators invoke \cite{Endl2002} as a justification. \cite{Endl2002} investigated the effect of eccentricity on their detection limit for HR~4979, using insertion and recovery. They test five planetary orbital periods $P_{\rm p}$, each with {\it one} corresponding semi-amplitude $K_{\rm p}$, and vary $e_{\rm p}$ over $9$ values and $\omega_{\rm p}$ over $4$, creating 180 $(P_{\rm p},K_{\rm p},e_{\rm p}, \omega_{\rm p})$ signals\footnote{strangely, \cite{Endl2002} only mention 110 signals}. They only test one phase angle $\phi_{\rm p}$.
At the time, producing these simulations represented a significant computational effort, however nowadays this appears as a rather coarse grid. From this exercise, \cite{Endl2002} conclude that eccentricity can affect detection limit calculations, but find that their detection limits assuming $e_{\rm p}=0$ are valid for $P_{\rm p} \gtrapprox 365 ~\rm days$, so long as the simulated Keplerian has $e_{\rm p}\lesssim0.5$. 
More recently \cite{Cumming2010} also concluded that assuming circular orbits provides a good agreement with upper semi-amplitude limits for $e_{\rm p}\lesssim0.5$ when recovering signals with a periodogram.
These particular results have been invoked to justify the assumption of circular orbits ever since.

Interestingly, for $P_{\rm p} \leq 365 ~\rm days$ \citep{Endl2002} find their detection limit is only valid if the inserted signal has $e<0.3$. However this recommendation has not been followed, and assuming circular orbits for short period is prevalent throughout the literature. From our analysis, we can show how eccentricity affects the detection limit over the entire period range. Our \kima runs contain $> 70,000$ posterior samples that naturally explore all orbital parameters.

\begin{figure}
	\includegraphics[width=\columnwidth]{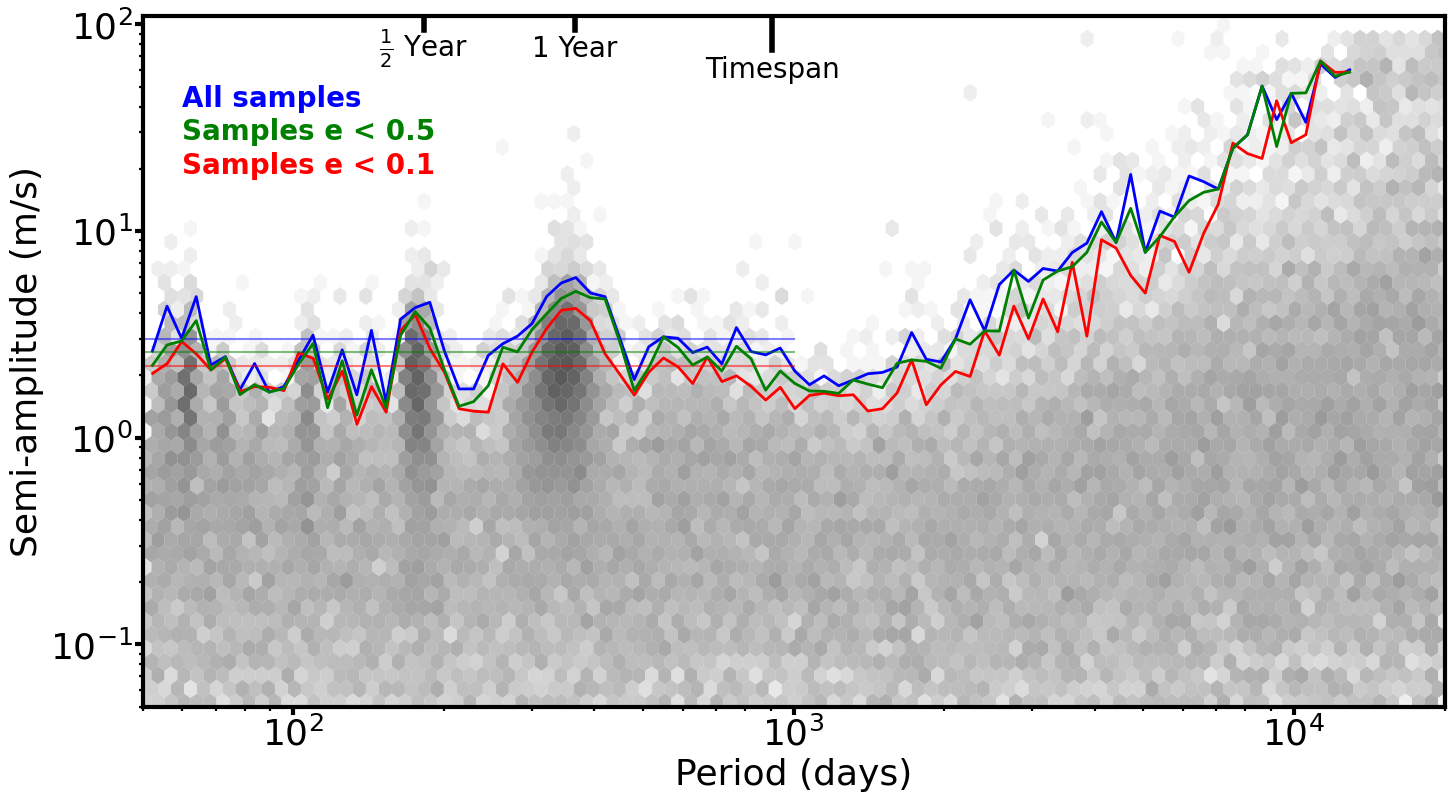}
    \caption{Hexbin plot of posterior samples and blue detection limit for J0310-31 as in figure~\ref{fig:J0310-31_Det_lims}. The Red and green lines here indicate the same limit calculated from posterior samples with eccentricities $<0.1$ and $<0.5$ respectively. Horizontal coloured lines show the mean semi-amplitude of the three coloured detection-limits out to $1000~\rm days$.}
    \label{fig:J0310-31_Det_lims_ecc}
\end{figure}

\begin{figure}
	\includegraphics[width=\columnwidth]{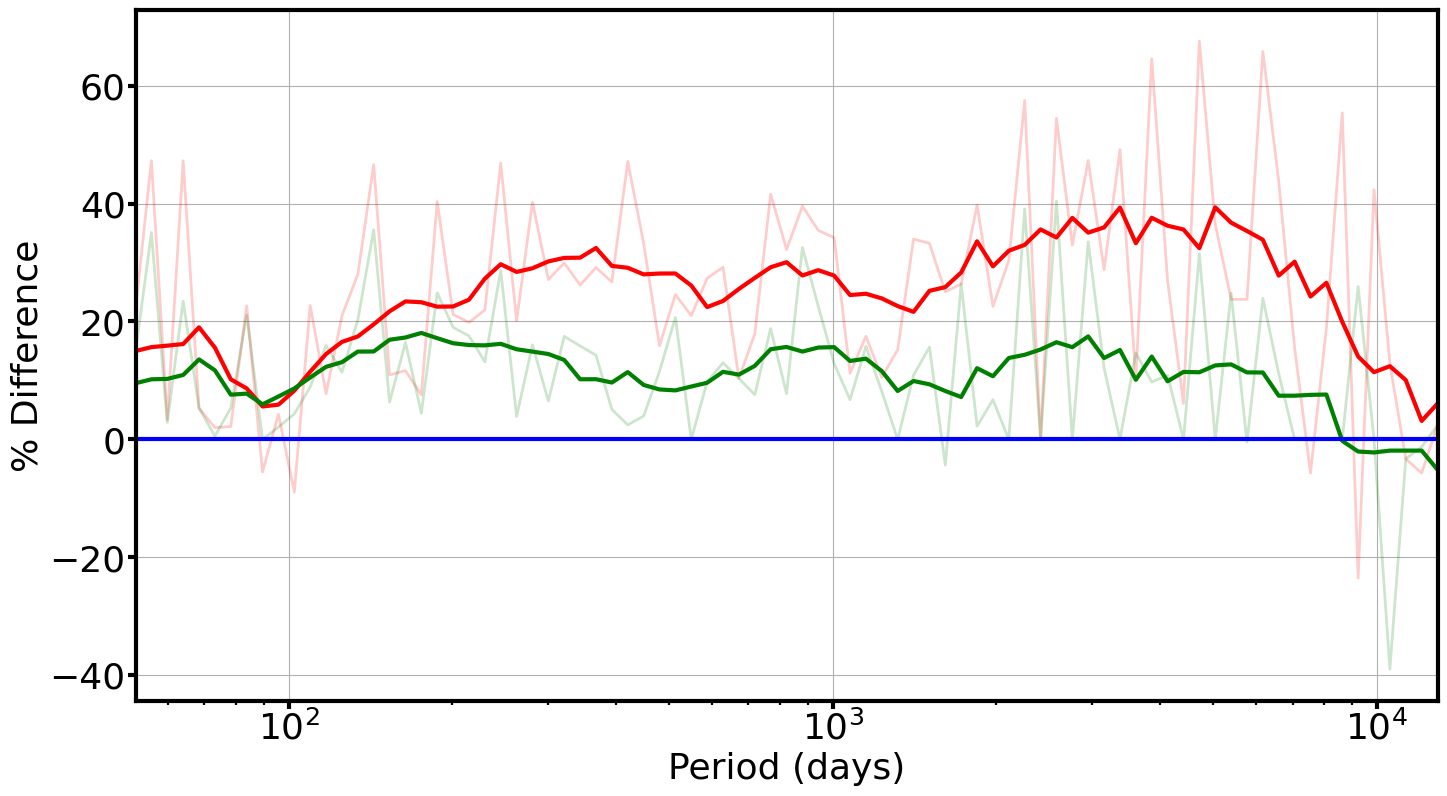}
    \caption{Plot of the percentage difference between detection limits calculated with all posterior samples, and posterior samples with eccentricities $<0.1$ and $<0.5$ in red and green respectively for J0310-31 from figure~\ref{fig:J0310-31_Det_lims_ecc}. Solid coloured lines represent a running mean taken on the faded coloured lines.}
    \label{fig:J0310-31_Det_lims_ecc_perc_diff}
\end{figure}

Figure~\ref{fig:J0310-31_Det_lims_ecc} shows the same detection limit for EBLM~J0310-31, as calculated before, in blue using all posterior samples. Alongside we plot a red detection limit produced in exactly the same way, except using only samples with $e_{\rm p} <0.1$, and green with $e_{\rm p} <0.5$ \citep[the limit stated by][]{Endl2002,Cumming2010}. For the three systems explored in this work, the red detection limit is systematically lower, by an average of $24.7\%$ or $\sim1~\rm m\,s^{-1}$ for periods $<1000~\rm day$. The green detection limit is also systematically lower by an average of $13.3\%$ or $\sim0.6~\rm m\,s^{-1}$ for the same periods.
Figure~\ref{fig:J0310-31_Det_lims_ecc_perc_diff} demonstrates how the two lower detection limits differ from the original for J0310-31 with period. We note an increase in divergence from the original detection limit at $P_{\rm p}$ exceeding the timespan of the data ($>1000~\rm day$ here), to a maximum of $39.9\%$ at $\sim 4000~\rm days$. This maximum difference corresponds to $3.5~\rm m\,s^{-1}$ ($\sim 120~\rm M_{\oplus}$)!
Our results are consistent with \cite{Wittenmyer2006} who are able to exclude planets with masses $\gtrsim 2~\rm M_{Jup}$ by assuming $e_{\rm p}=0$, and masses $\gtrsim 4~\rm M_{Jup}$ if instead they assume $e_{\rm p}=0.6$ at a given $P_{\rm p}\sim 4300 \rm ~days$, a $50\%$ difference, comparable to our $40\%$.

Here we remind the reader that we assume a Kumaraswamy distribution as our prior on $e_{\rm p}$, which already favours low $e_{\rm p}$, as is seen in observations \citep{Kipping2013}. Even though circumbinary planets discovered thus far are known to have relatively low eccentricities \citep[$e<0.15$][]{Martin2018}, we strongly caution against assuming circular orbits when calculating detection limits in any survey, particularly at long orbital periods.
Fixing $e_{\rm p}=0$ when calculating detection limits over-estimates the success of the program by a significant amount (up to $120~\rm M_{\oplus}$ or $0.4~\rm M_{Jup}$ in our case).

This exercise demonstrates the superiority of a diffusive nested sampler in establishing robust detection limits, which are crucial to infer the sensitivity of radial-velocity surveys, and consequently, the occurrence rates of exoplanets.

\section{Conclusions}\label{sec:conclusion}

\begin{figure*}
	\includegraphics[width=\textwidth]{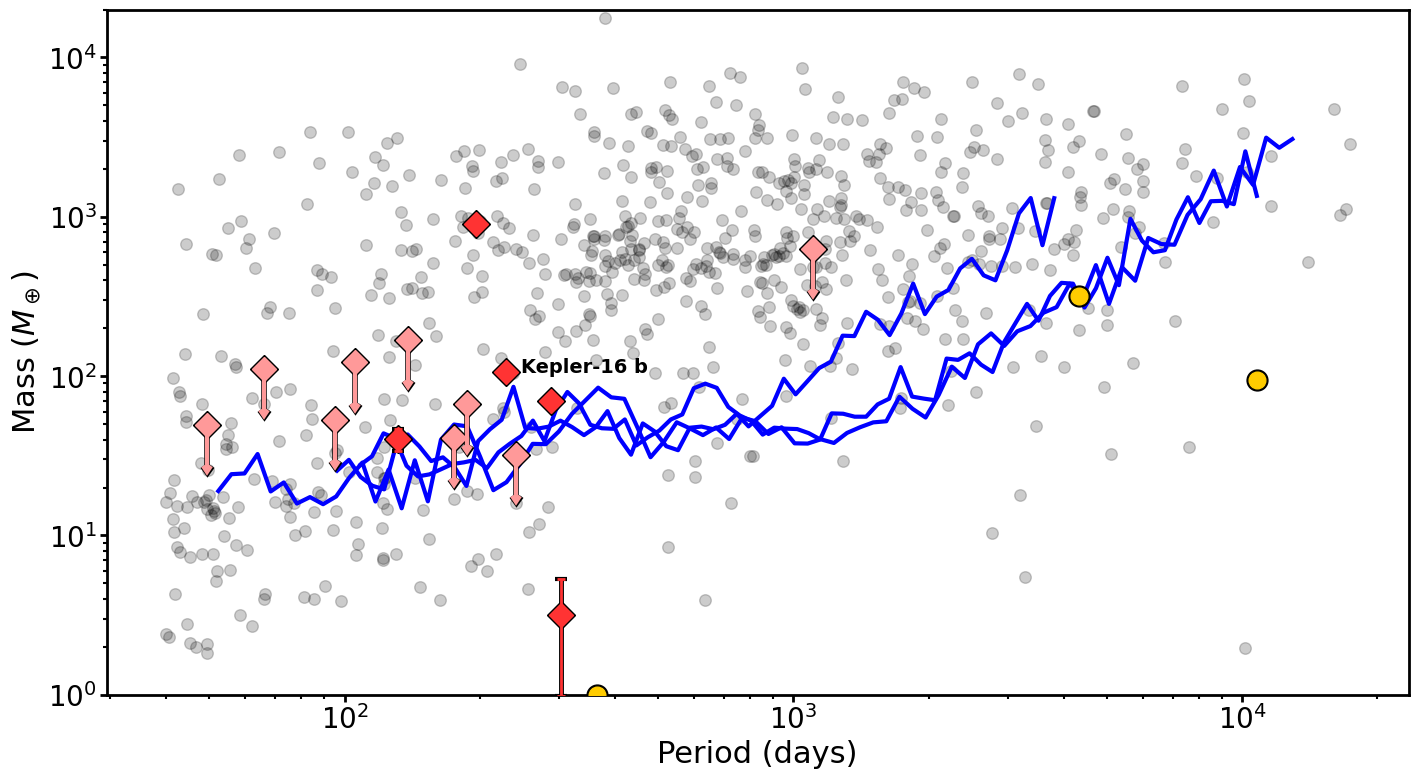}
    \caption{Mass vs period plot showing the three detection limits from this work as blue lines in comparison to confirmed exoplanets as grey circles, and transiting circumbinary planets as pink diamonds. Solar System planets are depicted as yellow dots for reference. Circumbinary planets in order of increasing period; \citetalias{Orosz2019}b, \citetalias{Kostov2014}, \citetalias{Kostov2020}, \citetalias{Orosz2012b}, \citetalias{Welsh2012}35~b, \citetalias{Schwamb2013}, \citetalias{Socia2020}, \citetalias{Orosz2019}d, \citetalias{Kostov2021}, \citetalias{Doyle2011}, \citetalias{Welsh2015}, \citetalias{Welsh2012}34~b, \citetalias{Orosz2019}c, \citetalias{Kostov2016}. Arrows illustrate planets with upper mass limits with the symbol placed to give a $2\sigma$ upper limit.}
    \label{fig:Detlims_mass_vs_P}
\end{figure*}

We analyse high-precision radial-velocities obtained as part of a large scale, ongoing, radial-velocity survey that seeks circumbinary planets using radial-velocities obtained with HARPS, ESPRESSO and SOPHIE, in both hemispheres, on single-lined eclipsing binaries. This survey is called BEBOP (Binaries Escorted By Orbiting Planets). We then detail an observing and Bayesian analysis protocol and test it on data collected for three single-lined binaries within the survey. Our analysis shows for the first time, a repeated ability to detect circumbinary planets with masses between Neptune's and Saturn's, for orbital periods within $1000$ days with as few as 25 spectra in the span of a year. Figure~\ref{fig:Detlims_mass_vs_P} displays a mass vs period plot of the detection limits from this work (blue) along with confirmed exoplanets (grey circles; NASA exoplanet archive\footnote{\url{https://exoplanetarchive.ipac.caltech.edu/index.html}}), transiting circumbinary planets  (magenta diamonds), and Solar System planets for comparison. This figure demonstrates our ability to detect sub-Saturn mass circumbinary planets at periods up to $1000$ days. Our data are able to detect a large fraction of currently known systems. We note though that many circumbinary planet detections made with eclipse timing variations are upper limits only.

We also present a method to compute detection limits on radial-velocity data by using a diffusive nested sampler, without the need to assume circular orbits as is the norm. We then show that this method is superior than the usual injection-recovery tests. 
Assuming circular orbits when determining detection limits generates over-optimistic detection limits by an average of $\sim 1~\rm{m\,s^{-1}}$ ($24.7\%$) at periods $<1000~\rm day$, and up to $120~\rm M_\oplus$ at periods $>1000~\rm day$. 
We therefore strongly caution against assuming circular orbits when calculating detection limits, and suggest \kima as a way for exoplanet surveys in general to extract accurate sensitivity limits and occurrence rates with fewer assumptions.

Thanks to the protocols and tests described in this paper, the BEBOP survey is now ready to produce circumbinary planet candidates following Bayesian evidence, and able to compute occurrence rates that can be compared to those already established from photometric methods \citep{Armstrong2014, Martin&Triaud2014}. Producing occurrence rates and upper limits on the occurrence of circumbinary planets will be done by simply combining all \kima produced detection limits.

\section*{Acknowledgements}
We would like to thank the staff at ESO's La Silla observatory for their hard work and assistance throughout this project, especially during the COVID pandemic. We also thank all of the observers who took part in the HARPS timeshare and were instrumental in collecting data for this projects. We particularly thank X. Dumusque and F. Bouchy for their work organising the timeshare.

This research received funding from the European Research Council (ERC) under the European Union’s Horizon 2020 research and innovation programme (grant agreement n$^{\circ}$803193/BEBOP) and from the Leverhulme Trust (research project n$^{\circ}$ RPG-2018-418). MRS would like to acknowledge the support of
the UK Science and Technology Facilities Council (STFC). We thank H. Rein for helpful discussion regarding this work. This work made use of the Astropy, numpy, pandas, scipy, corner, and matplotlib packages. We also use of \href{https://gist.github.com/lzkelley/0de9e8bf2a4fe96d2018f1b1bd5a0d3}{label$\_$line.py} by Luke Zoltan Kelley. The French group acknowledges financial support from the French Programme National de Plan\'etologie (PNP, INSU). 
ACMC acknowledges support by CFisUC projects (UIDB/04564/2020 and UIDP/04564/2020), GRAVITY (PTDC/FIS-AST/7002/2020), ENGAGE SKA (POCI-01-0145-FEDER-022217), and PHOBOS (POCI-01-0145-FEDER-029932), funded by COMPETE 2020 and FCT, Portugal.

\section*{Data Availability}

The data we used is available for download at the ESO public archive\footnote{\url{http://archive.eso.org/eso/eso_archive_main.html}}. Search for Prog.ID 099.C-0138 and 1101.C-0721.



\bibliographystyle{mnras}
\bibliography{Circumbinary_bib} 




\appendix

\section{Injection and Recovery results and Radial Velocity data}

Here we present the results of our insertion/recovery tests for each of our three target systems in tables~\ref{tab:0310_injrec} to \ref{tab:1928_injrec}, Numbers in the tables are Bayes factors calculated from our \kima runs. We also present our RV data on the systems in tables~\ref{tab:J0310-31_data} to \ref{tab:J1928-38_data}.

\begin{table*}
	\centering
	\caption{J0310-31 Keplerian insertion and recovery results}
	\label{tab:0310_injrec}
	\begin{tabular}{l|cccccc}
		\hline
		 Semi-amplitudes & & & Periods & &\\
		 & $5.9\times P_{\rm bin} = 74.59 \rm{d}$ & $12\times P_{\rm bin} = 151.71 \rm{d}$ & $18\times P_{\rm bin} = 227.57 \rm{d}$ & $30\times P_{\rm bin} = 379.28 \rm{d}$ & $486.67 \rm{d}$ & $400\times P_{\rm bin} = 5057.12 \rm{d}$\\
		\hline
		$10\times \rm{RMS} = 31.6\rm{m\,s^{-1}}$ & - & - & - & - & - & inf\\
		$3\times \rm{RMS} = 9.48\rm{m\,s^{-1}}$ & - & - & - & inf & - & $8.3$\\
		$2\times \rm{RMS} = 6.32\rm{m\,s^{-1}}$ & inf & - & - & - & - & -\\
		$1.5\times \rm{RMS} = 4.74\rm{m\,s^{-1}}$ & $39.7$ & inf & - & $25.7$ & $5946$ & $0.8$\\
		$1.25\times \rm{RMS} = 3.95\rm{m\,s^{-1}}$ & $3.2$ & $5860$ & $984.83$ & - & $409.2$ & -\\
		$\rm{RMS} = 3.16\rm{m\,s^{-1}}$ & - & $1070.7$ & $120.6$ & $0.9$ & $44.7$ & -\\
		$0.75\times \rm{RMS} = 2.37\rm{m\,s^{-1}}$ & - & $63.7$ & $16.4$ & - & $7.6$ & -\\
		$0.5\times \rm{RMS} = 1.58\rm{m\,s^{-1}}$ & - & $5.8$ & $2.5$ & - & - & -\\
		\hline
	\end{tabular}
\end{table*}

\begin{table*}
	\centering
	\caption{J0310-31 Rebound simulated N-body insertion and recovery results}
	\label{tab:0310_injrec_rebound}
	\begin{tabular}{l|ccc}
		\hline
		 Semi-amplitudes & & Periods &\\
		 & $5.9\times P_{\rm bin} = 74.59 \rm{d}$ & $12\times P_{\rm bin} = 151.71 \rm{d}$ & $486.67 \rm{d}$\\
		\hline
		$2\times \rm{RMS} = 6.32\rm{m\,s^{-1}}$ & inf & - & -\\
		$1.5\times \rm{RMS} = 4.74\rm{m\,s^{-1}}$ & $980$ & inf & $1378.8$ \\
		$1.25\times \rm{RMS} = 3.95\rm{m\,s^{-1}}$ & $60.5$ & inf & $264.5$\\
		$\rm{RMS} = 3.16\rm{m\,s^{-1}}$ & $5.2$ & $7020$ & $41.5$\\
		$0.75\times \rm{RMS} = 2.37\rm{m\,s^{-1}}$ & - & $217.4$ & $7$\\
		$0.5\times \rm{RMS} = 1.58\rm{m\,s^{-1}}$ & - & $11.3$ & -\\
		\hline
	\end{tabular}
\end{table*}

\begin{table*}
	\centering
	\caption{J1540-09 Keplerian insertion and recovery results}
	\label{tab:1540_injrec}
	\begin{tabular}{l|cccc}
		\hline
		 Semi-amplitudes & & Periods & &\\
		 & $5.9\times P_{\rm bin} = 155.4 \rm{d}$ & $12\times P_{\rm bin} = 316.06 \rm{d}$ & $486.67 \rm{d}$ & $999 \rm{d}$\\
		\hline
		$3\times \rm{RMS} = 11.25\rm{m\,s^{-1}}$ & - & inf & - & -\\
		$2.5\times \rm{RMS} = 9.38\rm{m\,s^{-1}}$ & - & inf & - & -\\
		$2\times \rm{RMS} = 7.5\rm{m\,s^{-1}}$ & inf & $637.3$ & inf & inf\\
		$1.5\times \rm{RMS} = 5.63\rm{m\,s^{-1}}$ & $2811.7$ & $14.5$ & $1276.8$ & $291$\\
		$1.25\times \rm{RMS} = 4.69\rm{m\,s^{-1}}$ & $464.5$ & $3.6$ & $223.2$ & $29.5$\\
		$\rm{RMS} = 3.75\rm{m\,s^{-1}}$ & $72.6$ & - & $23.2$ & $3.7$\\
		$0.75\times \rm{RMS} = 2.81\rm{m\,s^{-1}}$ & $8.4$ & - & $3.6$ & -\\
		$0.5\times \rm{RMS} = 1.88\rm{m\,s^{-1}}$ & $1.5$ & - & $0.9$ & $0.7$\\
		\hline
	\end{tabular}
\end{table*}

\begin{table*}
	\centering
	\caption{J1928-38 Keplerian insertion and recovery results}
	\label{tab:1928_injrec}
	\begin{tabular}{l|ccccc}
		\hline
		 Semi-amplitudes & & & Periods &\\
		 & $5.1\times P_{\rm bin} = 118.95 \rm{d}$ & $5.9\times P_{\rm bin} = 137.61 \rm{d}$ & $8\times P_{\rm bin} = 186.58 \rm{d}$ & $12\times P_{\rm bin} = 279.87 \rm{d}$ & $486.67 \rm{d}$\\
		\hline
		$5\times \rm{RMS} = 14.8\rm{m\,s^{-1}}$ & - & - & - & - & inf\\
		$4\times \rm{RMS} = 11.84\rm{m\,s^{-1}}$ & - & inf & - & inf & inf\\
		$3\times \rm{RMS} = 8.88\rm{m\,s^{-1}}$ & $2919$ & $1362.6$ & $6413$ & inf & inf\\
		$2\times \rm{RMS} = 5.92\rm{m\,s^{-1}}$ & $49.8$ & $21$ & $47.8$ & $58.9$ & $138.4$\\
		$1.5\times \rm{RMS} = 4.44\rm{m\,s^{-1}}$ & - & $2.4$ & - & $2.9$ & $14.3$\\
		$1.25\times \rm{RMS} = 3.7\rm{m\,s^{-1}}$ & $2.3$ & - & $2.9$ & - & -\\
		$\rm{RMS} = 2.96\rm{m\,s^{-1}}$ & - & - & - & - & $2.4$\\
		\hline
	\end{tabular}
\end{table*}

\begin{table*}
	\centering
	\caption{Journal of Observations containing our HARPS data for J0310-31. Dates are given in BJD - 2,400,000. $V_{\rm rad}$ are the measured radial velocities with their uncertainties $\sigma_{V_{\rm rad}}$. FWHM is the Full With at Half Maximum of the Gaussian fitted to the cross correlation function, and {\it contrast} is its amplitude. bis. span is the span of the bisector slope. Uncertainties on FWHM and bis. span are $2\times\sigma_{V_{\rm rad}}$.}
	\label{tab:J0310-31_data}
	\begin{tabular}{llllll} 
		\hline
		\hline
        BJD&$V_{\rm rad}$&$\sigma_{V_{\rm rad}}$&FWHM&contrast&bis. span\\
	&[km\,s$^{-1}$]&[km\,s$^{-1}$]&[km\,s$^{-1}$]& &[km\,s$^{-1}$]\\
		\hline
57943.892368&35.63948&0.00168&10.93301&31.049&0.02153\\
57944.923580&24.07411&0.00170&10.95037&31.088&0.02859\\
57945.885900&8.58139&0.00176&10.92966&31.091&0.01777\\
57946.876879&-6.07696&0.00185&10.92368&31.127&0.02400\\
57948.928854&18.35127&0.00199&10.92909&31.150&0.02186\\
57949.874496&32.36601&0.00376&10.93177&31.081&0.00910\\
57951.890174&46.34736&0.00245&10.92515&31.176&0.02321\\
57952.924664&48.32340&0.00184&10.93519&31.078&0.02634\\
57953.919672&47.93648&0.00224&10.93547&31.070&0.03302\\
57955.935525&40.23726&0.00148&10.92350&31.114&0.02441\\
57956.937097&31.73699&0.00242&10.90176&31.215&0.02218\\
57958.914371&1.85626&0.00250&10.90546&31.207&0.02907\\
57959.923988&-7.18919&0.00185&10.91742&31.126&0.02583\\
57960.925502&6.01245&0.00218&10.90108&31.196&0.01583\\
57962.941783&36.84076&0.00240&10.91809&31.103&0.02995\\
57967.873028&44.11773&0.00509&10.91773&31.108&0.03909\\
57968.862759&38.22158&0.00196&10.93156&31.078&0.02316\\
57970.834614&14.43370&0.00201&10.93765&31.107&0.02358\\
57971.887207&-3.14864&0.00203&10.90942&31.205&0.02815\\
57972.859262&-5.30070&0.00212&10.91962&31.137&0.02385\\
57973.874664&11.99910&0.00246&10.95580&31.021&0.03180\\
57974.855492&28.50055&0.00207&10.94437&31.112&0.01107\\
57979.910020&46.38457&0.00172&10.92281&31.125&0.03514\\
57980.907132&42.15442&0.00189&10.90393&31.195&0.03216\\
57981.912450&34.81087&0.00166&10.91749&31.142&0.02222\\
57993.897684&40.01640&0.00175&10.89351&31.237&0.02857\\
57994.868121&31.71289&0.00204&10.90854&31.176&0.01557\\
57995.913932&17.82943&0.00210&10.96594&31.005&0.02541\\
57996.900885&0.89748&0.00326&10.95410&31.062&0.01592\\
57998.846013&5.86085&0.00207&10.93080&31.128&0.03219\\
58000.814335&36.31122&0.00184&10.93406&31.090&0.02512\\
58001.825348&43.67481&0.00173&10.93076&31.065&0.03359\\
58002.875888&47.45290&0.00195&10.90752&31.138&0.02670\\
58008.865205&12.69217&0.00302&10.91455&31.160&0.02339\\
58010.891199&-4.13140&0.00193&10.93348&31.103&0.02658\\
58011.853533&12.96220&0.00186&10.93126&31.120&0.02981\\
58012.842881&29.29996&0.00194&10.93123&31.138&0.02439\\
58013.836117&39.60231&0.00231&10.89094&31.251&0.03815\\
58021.844325&6.77414&0.00237&10.88521&31.277&0.02190\\
58022.865563&-6.90867&0.00248&10.90068&31.182&0.02454\\
58023.861648&0.85340&0.00259&10.87215&31.286&0.01533\\
58026.843370&42.13524&0.00420&10.85961&31.323&0.03642\\
58027.851246&46.66027&0.00274&10.96376&30.975&0.03885\\
58292.935003&45.19779&0.00188&10.85673&31.290&0.03447\\
58295.939322&46.50750&0.00153&10.87040&31.248&0.02880\\
58336.906794&23.28586&0.00177&10.85126&31.372&0.01926\\
58338.851019&-6.42041&0.00176&10.85513&31.308&0.02670\\
58354.838158&36.55513&0.00227&10.87155&31.222&0.02979\\
58405.803761&39.90978&0.00460&10.85632&31.338&0.02138\\
58453.710487&0.74084&0.00138&10.90947&31.141&0.02863\\
58488.633260&23.11336&0.00169&10.86625&31.319&0.03055\\
58506.584524&36.86690&0.00139&10.91340&31.117&0.02021\\
58515.547144&-3.39784&0.00178&10.93431&31.069&0.02643\\
58519.564335&39.72933&0.00178&10.90851&31.130&0.02260\\
58692.848790&-6.41913&0.00154&10.90509&31.272&0.03065\\
58701.837991&40.38147&0.00167&10.91534&31.217&0.02113\\
58707.811668&23.67250&0.00297&10.91707&31.212&0.02879\\
58723.756754&47.90169&0.00143&10.91435&31.183&0.02889\\
58727.806796&35.08783&0.00171&10.91437&31.193&0.02715\\
58753.740711&28.24276&0.00219&10.92187&31.189&0.02651\\
58762.749317&48.28098&0.00163&10.91574&31.195&0.03644\\
58784.717703&36.78734&0.00202&10.89169&31.348&0.01632\\
58834.649676&29.63072&0.00147&10.95468&31.105&0.02506\\
58846.646146&19.74657&0.00175&10.95264&31.044&0.02730\\
58850.565260&48.32348&0.00127&10.92443&31.170&0.03143\\
		\hline
	\end{tabular}
\end{table*}

\begin{table*}
	\centering
	\caption{Journal of Observations containing our HARPS data for J1540-09. Columns are as in \ref{tab:J0310-31_data}}
	\label{tab:J1540-09_data}
	\begin{tabular}{llllll} 
		\hline
		\hline
        BJD&$V_{\rm rad}$&$\sigma_{V_{\rm rad}}$&FWHM&contrast&bis. span\\						&[km\,s$^{-1}$]&[km\,s$^{-1}$]&[km\,s$^{-1}$]& &[km\,s$^{-1}$]\\
		\hline
57863.886017&-34.04435&0.00284&8.28287&34.515&0.03306\\
57865.839152&-43.91032&0.00232&8.29807&34.391&0.02892\\
57872.835002&-77.11478&0.00297&8.27285&34.583&0.02351\\
57875.857985&-74.41238&0.00535&8.29200&34.511&0.01455\\
57879.754279&-61.45253&0.00319&8.28987&34.441&0.03454\\
57923.573674&-73.41843&0.00409&8.25559&34.767&0.03070\\
57934.687091&-51.51693&0.00429&8.25721&34.747&0.03096\\
57942.695102&-33.38362&0.00351&8.30337&34.065&0.03405\\
57944.691282&-42.89582&0.00281&8.31499&34.146&0.02189\\
57945.660293&-49.29063&0.00272&8.30231&34.259&0.03662\\
57948.624456&-68.11507&0.00343&8.28781&34.553&0.01667\\
57951.646824&-76.94824&0.00745&8.26195&34.595&0.03799\\
57959.584739&-57.97578&0.00382&8.28518&34.608&0.03248\\
57964.618682&-36.33938&0.00429&8.31324&34.080&0.04210\\
57969.547587&-35.17730&0.00473&8.27743&34.448&0.04891\\
57979.581918&-76.95182&0.00272&8.29949&34.246&0.02375\\
57996.529560&-38.13035&0.00324&8.31348&34.277&0.04163\\
58001.510009&-69.13364&0.00302&8.30714&34.214&0.02582\\
58008.470052&-72.09597&0.02062&8.26838&34.398&0.07094\\
58012.493882&-56.95128&0.00442&8.28094&34.427&0.01261\\
58258.703952&-33.23617&0.00278&8.24429&34.880&0.03156\\
58272.646623&-69.65691&0.00254&8.25571&34.828&0.02267\\
58276.707794&-53.23456&0.00213&8.27952&34.650&0.02728\\
58278.649982&-44.46367&0.00270&8.26452&34.718&0.01897\\
58286.693559&-40.52268&0.00279&8.25444&34.814&0.04001\\
58291.571082&-70.64953&0.00291&8.26216&34.799&0.03911\\
58301.682725&-59.27470&0.00501&8.26292&34.601&0.04623\\
58334.515634&-32.95173&0.00224&8.26416&34.732&0.03756\\
58340.485380&-47.58266&0.00313&8.23341&34.979&0.02739\\
58356.492448&-49.72878&0.00241&8.25519&34.600&0.02249\\
58521.884427&-32.68384&0.00572&8.24942&34.789&0.03390\\
58548.884978&-34.78751&0.00269&8.27384&34.712&0.02978\\
58565.835269&-55.89578&0.00223&8.29513&34.526&0.02256\\
58584.849299&-77.25342&0.00212&8.30491&34.430&0.01920\\
58600.898483&-32.67798&0.00320&8.32099&34.343&0.03645\\
58611.764016&-76.90855&0.00584&8.28568&34.680&0.02539\\
58615.819016&-66.97010&0.00252&8.29986&34.518&0.01559\\
58644.693557&-56.58981&0.00235&8.29114&34.526&0.03632\\
58660.683192&-72.15472&0.00377&8.27641&34.651&0.03250\\
58673.679701&-44.66300&0.00441&8.27281&34.705&0.01648\\
58914.890985&-31.32791&0.00276&8.27327&34.612&0.03724\\
		\hline
	\end{tabular}
\end{table*}

\begin{table*}
	\centering
	\caption{Journal of Observations containing our HARPS data for J1928-38. Columns are as in \ref{tab:J0310-31_data}}
	\label{tab:J1928-38_data}
	\begin{tabular}{llllll} 
		\hline
		\hline
        BJD&$V_{\rm rad}$&$\sigma_{V_{\rm rad}}$&FWHM&contrast&bis. span\\
&[km\,s$^{-1}$]&[km\,s$^{-1}$]&[km\,s$^{-1}$]& &[km\,s$^{-1}$]\\
		\hline
58273.834134&32.80650&0.00217&7.04698&48.100&-0.00888\\
58278.843949&17.69958&0.00355&7.07692&47.503&-0.00764\\
58284.752885&-1.49688&0.00232&7.04283&48.268&-0.00407\\
58288.745112&9.82687&0.00190&7.04846&48.270&-0.00610\\
58292.729051&25.70995&0.00326&7.02329&48.428&0.00265\\
58293.708035&28.62785&0.00290&7.03782&47.844&0.00469\\
58311.790430&8.67845&0.00209&7.05597&47.813&-0.00506\\
58334.565596&6.49932&0.00184&7.05307&48.138&0.00278\\
58338.621368&23.08744&0.00361&7.03939&48.057&-0.00714\\
58346.585986&27.43032&0.00176&7.05339&48.206&-0.01117\\
58355.645102&-0.30616&0.00228&7.06381&47.595&-0.00554\\
58550.896768&30.15923&0.00242&7.07663&47.847&-0.00196\\
58556.881115&26.03178&0.00374&7.06599&48.034&0.00072\\
58556.901788&25.94663&0.00338&7.06541&48.122&-0.01728\\
58564.890277&-1.28170&0.00189&7.06720&47.828&-0.00538\\
58586.899133&-1.30374&0.00203&7.07108&47.947&-0.00910\\
58612.855107&1.20469&0.00577&7.07242&47.921&-0.01711\\
58637.829855&6.75596&0.00257&7.07361&47.829&-0.00782\\
58643.837284&29.35475&0.00212&7.09607&47.618&0.00213\\
58664.803891&21.81645&0.00247&7.11147&47.332&0.00270\\
58667.703897&30.57137&0.00231&7.08125&47.868&0.00740\\
58670.740859&32.53833&0.00203&7.07063&47.896&-0.00829\\
58700.663044&7.52499&0.00187&7.07861&47.860&-0.00283\\
58726.538304&-0.92379&0.00392&7.05509&47.966&-0.00253\\
58740.520279&32.68685&0.00267&7.06985&47.567&-0.01475\\
		\hline
	\end{tabular}
\end{table*}


\bsp	
\label{lastpage}
\end{document}